\documentclass[a4paper,12pt, epsfig]{article}
\def\letter{0}\def\pr{0}
\pdfoutput=1 
\usepackage{epsfig}
\usepackage{epstopdf}
\usepackage{graphicx}
\usepackage{ifthen}
\usepackage{ulem}

\usepackage{amsmath}
\usepackage[psamsfonts]{amssymb}
\usepackage{euscript}

\usepackage{latexsym}
\usepackage[arrow,matrix,curve]{xy}

\usepackage[hypertexnames=false]{hyperref}
\hypersetup{
    colorlinks=true,
    linkcolor=blue,
    filecolor=magenta,   
    citecolor=black,   
    urlcolor=cyan,
    }

\newskip\humongous \humongous=0pt plus 1000pt minus 1000pt

\newif\ifdtup

\def\,{\hspace{-.1cm}}
\def\hsp{,\hspace{.7cm}}

\def\fc#1#2 {\frac{n}{q}#1\frac{n}{q}#2}

\def\kt{\kappa}
\def\kt{\mathfrak{K}}

\def\hpt{H_{\rm{free}}}

\newcommand{\vac}{\ensuremath{|0\rangle}}

\renewcommand{\sin}{\textrm{sin}}

\renewcommand{\sinh}{\textrm{sinh}}

\renewcommand{\tanh}{\textrm{tanh}}
\newcommand{\sech}{\textrm{sech}}
\newcommand{\csch}{\textrm{csch}}

\def\sl{{\sqrt{\lambda}}}

\renewcommand{\theequation}{\arabic{section}.\arabic{equation}}
\renewcommand{\(}{\begin{equation}}
\renewcommand{\)}{end{equation} \vspace{-.05in}\linebreak}

\newcounter{saveeqn}
\newcounter{savealpheqn}

\newcommand{\alpheqn}{\setcounter{saveeqn}{\value{equation}}%
  \stepcounter{saveeqn}\setcounter{equation}{0}%
  \renewcommand{\theequation}{\mbox{\arabic{section}.\arabic{saveeqn}
\alph{equation}}}
  \renewcommand{\)}{\end{equation}}}
\def\part#1{\frac{\partial}{\partial{#1}}}%
\def\group#1{\refstepcounter{equation}\setcounter{saveeqn}
 {\value{equation}}%
  \label{#1}\setcounter{equation}{0}%
\renewcommand{\theequation}{\mbox{\arabic{section}.\arabic{saveeqn}
\alph{equation}}}
  \renewcommand{\)}{\end{equation}}}
\newcommand{\reseteqn}{\setcounter{equation}{\value{saveeqn}}%
  \renewcommand{\theequation}{\arabic{section}.\arabic{equation}}%
  \renewcommand{\)}{\end{equation}}}

\newcommand{\aalpheqn}{\setcounter{saveeqn}{\value{equation}}%
  \stepcounter{saveeqn}\setcounter{equation}{0}%
  \renewcommand{\theequation}{\mbox{
        \Alph{subsection}.\arabic{saveeqn}\alph{equation}}}
   \renewcommand{\)}{\end{equation}}}
\newcommand{\areseteqn}{\setcounter{equation}{\value{saveeqn}}%
  \renewcommand{\theequation}{\Alph{subsection}.\arabic{equation}}%
  \renewcommand{\)}{\end{equation}}}

\renewcommand{\thefootnote}{\alph{footnote}}
\renewcommand{\(}{\begin{equation}}
\renewcommand{\)}{\end{equation}}
\newcommand{\ba}{\begin{eqnarray}}
\newcommand{\ea}{\end{eqnarray}}

\renewcommand{\b}{\beta}

\newcommand{\cbp}{\mathop{\vtop{\ialign{##\crcr
   $\hfil\displaystyle{}\hfil$\crcr\noalign{\kern-13pt\nointerlineskip}
   \BIG{)}\hskip0pt\crcr\noalign{\kern3pt}}}}}
\newcommand{\pa}{\mathop{\vtop{\ialign{##\crcr

$\hfil\displaystyle{\oplus}\hfil$\crcr\noalign{\kern+1pt\nointerlineskip
}
   \hspace{.08in}$^{\alpha=0}$\hskip6pt\crcr\noalign{\kern3pt}}}}}
\renewcommand{\hsp}{,\hspace{.3in}}
\newcommand{\p}{^\prime}

\catcode`\@=11
\def\vereq#1#2{\lower3pt\vbox{\baselineskip1.5pt \lineskip1.5pt
\ialign{$\m@th#1\hfill##\hfil$\crcr#2\crcr\sim\crcr}}}
\catcode`\@=12

\renewcommand{\(}{\begin{equation}}
\renewcommand{\)}{\end{equation}}

\def\pin#1{\int \frac{d#1}{2\pi}}
\def\ppin#1{\int\hspace{-17pt}\sum \frac{d#1}{2\pi}}
\def\ppink#1{\int\hspace{-17pt}\sum\frac{d^{#1}k}{(2\pi)^{#1}}}

\def\dint{\int\hspace{-12pt}\sum }
\def\pink#1{\int \frac{d^{#1}k}{(2\pi)^{#1}}}

\def\cc{\mathcal{C}}
\def\df{\mathcal{D}_{f}}

\def\blu#1{\textcolor{magenta}{Jarah: #1}}
\def\red#1{\textcolor{red}{Hui: #1}}
\def\corr#1{{#1}}

\newcommand{\beas}{\begin{eqnarray*}}
\newcommand{\eeas}{\end{eqnarray*}}

\newcommand{\bquo}{\begin{quote}}
\newcommand{\enqu}{\end{quote}}

\newcommand{\R}{{\mathbb R}}

\newcommand{\g}{{\mathfrak g}}

\def\ok#1{\omega_{k_{#1}}}
\def\okp#1{\omega_{k\p_{#1}}}
\def\V#1{V^{(#1)}(\sqrt{\lambda}f(x))}

\def\ck{\csch\left(\frac{\pi k}{2\b}\right)}

\def\mb{\mathcal{B}}
\def\mc{\mathcal{C}}
\def\md{\mathcal{D}}
\def\me{\mathcal{E}}

\newcommand{\beq}{\begin{equation}}
\newcommand{\eeq}{\end{equation}}
\newcommand{\bea}{\begin{eqnarray}}
\newcommand{\eea}{\end{eqnarray}}

\newskip\humongous \humongous=0pt plus 1000pt minus 1000pt

\newif\ifdtup

\catcode`\@=11

\ifthenelse{\equal{\letter}{0}}{ 

\@addtoreset{equation}{section}
\def\theequation{\arabic{section}.\arabic{equation}}

\def\@normalsize{\@setsize\normalsize{15pt}\xiipt\@xiipt
\abovedisplayskip 14pt plus3pt minus3pt%
\belowdisplayskip \abovedisplayskip
\abovedisplayshortskip \z@ plus3pt%
\belowdisplayshortskip 7pt plus3.5pt minus0pt}

\def\small{\@setsize\small{13.6pt}\xipt\@xipt
\abovedisplayskip 13pt plus3pt minus3pt%
\belowdisplayskip \abovedisplayskip
\abovedisplayshortskip \z@ plus3pt%
\belowdisplayshortskip 7pt plus3.5pt minus0pt
\def\@listi{\parsep 4.5pt plus 2pt minus 1pt
      \itemsep \parsep
      \topsep 9pt plus 3pt minus 3pt}}

\relax

\def\section{\@startsection{section}{1}{\z@}{3.5ex plus 1ex minus  .2ex}{2.3ex plus .2ex}{\large\bf}}

\def\thesection{\arabic{section}}
\def\thesubsection{\arabic{section}.\arabic{subsection}}

\def\appendix{\setcounter{section}{0}
 \def\thesection{Appendix \Alph{section}}
 \def\thesubsection{\Alph{section}.\arabic{subsection}}
 \def\theequation{\Alph{section}.\arabic{equation}}}
\renewcommand{\theequation}{\arabic{section}.\arabic{equation}}

}{
\renewcommand{\theequation}{\arabic{equation}}

} 

\begin{document}
\def\thefootnote{\fnsymbol{footnote}}
\def\thetitle{Kinks Multiply Mesons}
\def\autone{Hui Liu}
\def\auttwo{Jarah Evslin}
\def\autthree{Baiyang Zhang}
\def\affa{School of Fundamental Physics and Mathematical Sciences, Hangzhou Institute for Advanced Study,
University of Chinese Academy of Sciences, Hangzhou 310024, China}
\def\affb{University of the Chinese Academy of Sciences, YuQuanLu 19A, Beijing 100049, China}
\def\affc{Arnold Sommerfeld Center, Ludwig-Maximilians-Universität, Theresienstraße 37, 80333 München, Germany}
\def\affd{Institute of Modern Physics, NanChangLu 509, Lanzhou 730000, China}
\def\affe{Institute of Contemporary Mathematics, School of Mathematics and Statistics,
Henan University, Kaifeng, Henan 475004, P. R. China}


\ifthenelse{\equal{\pr}{1}}{
\title{\thetitle}
\author{\autone}
\author{\auttwo}
\author{\autthree}
\affiliation {\affa}
\affiliation {\affb}

}{

\begin{center}
{\large {\bf \thetitle}}

\bigskip

\bigskip


{\large \noindent  \autone{${}^{1,2,3}$} \footnote{hui.liu@campus.lmu.de} 
, \auttwo{${}^{4,2}$} \footnote{jarah@impcas.ac.cn}
and \autthree{${}^{5}$} \footnote{byzhang@henu.edu.cn}}


\vskip.7cm

1) \affa\\
2) \affb\\
3) \affc\\
4) \affd\\
5) \affe\\

\end{center}

}

\begin{abstract}
\noindent
In a (1+1)-dimensional scalar quantum field theory, we calculate the leading-order probability of meson multiplication, which is the inelastic scattering process: kink + meson $\rightarrow$ kink + 2 mesons.  We also calculate the differential probability with respect to the final meson momenta and the probability that one or two of the final mesons recoils back towards the source.  In the ultrarelativistic limit of the initial meson, the total probability tends to a constant, which we calculate analytically in the $\phi^4$ model.  At this order the meson sector conserves energy on its own, while the incoming meson applies a positive pressure to the kink.  This is in contrast with the situation in classical field theory, where Romanczukiewicz and collaborators have shown that, in the presence of a reflectionless kink, only meson fusion is allowed, resulting in a negative radiation pressure on the kink.

\end{abstract}

%
\setcounter{footnote}{0}
\renewcommand{\thefootnote}{\arabic{footnote}}

\ifthenelse{\equal{\pr}{1}}
{
\maketitle
}{}

\section{Introduction}
Two-dimensional scalar models provide an ideal sandbox for developing tools to treat real-world solitons.  If a scalar field is subjected to a potential with degenerate minima, then the theory will enjoy kink and antikink solutions.  In general, at weak coupling, one can decompose a given configuration into kinks and also perturbative, elementary quanta of the scalar field, called mesons.  An understanding of these theories at weak coupling is then reduced to understanding the interactions of mesons with one another, of kinks with (anti)kinks and of kinks with mesons.

The interactions of mesons with one another is largely as in the perturbative theory with no kinks, and so is well understood.  Interactions of kinks with (anti)kinks in classical field theory is a rich field and has been a subject of intense investigation since the discovery of resonance windows \cite{csw} and related phenomena \cite{osc,osc3d}.  It was once thought that these phenomena can be understood in terms of the internal excitations of the kink, but it has been found in Ref.~\cite{doreyf6} that resonances persist in the $\phi^6$ theory, whose kink has no internal excitations.  Instead, although certainly the internal excitations do affect the scattering phenomenology \cite{multex22a,multex22b}, it is now widely believed \cite{sfal21,col22} that a decisive role is played by the interactions of kinks with bulk excitations, which are not localized to a single kink and in this sense are related to mesons.

Kink-meson interactions have received relatively little attention, despite being the simplest nonperturbative scattering processes in such models.  In classical field theory, the mesons correspond to radiation.  Using the perturbative approach to the classical equations of motion for radiation introduced in Ref.~\cite{mm}, incident radiation upon a kink was studied in Refs.~\cite{tomrad1,tomrad2}.  It was found that if the kink is reflectionless, and the radiation is monochromatic with frequency $\omega$, then some of the transmitted radiation will have a frequency of $2\omega$ and this frequency doubling will exert a negative pressure on the kink.  In a quantized model this is easy to understand, it represents the process kink$+2$mesons$ \rightarrow $kink$+$meson.  One can show that energy conservation among the mesons, which is exact at leading order, implies that the final state meson has more momentum than the two merged mesons, with the difference causing a negative recoil of the kink.  This, including higher-order meson merging, is the only processes admitted in the case of classical reflectionless kinks.  In the case of reflective kinks, Ref.~\cite{tomrad3} found that there is also meson reflection, yielding a positive contribution to the pressure.

In the present note we consider a new process, meson multiplication, in which a meson incident on a kink splits into two mesons.  This process appears to have no classical counterpart, in the sense that the perturbative approach of Ref.~\cite{mm} is able to solve any initial value problem which begins with frequency $\omega$ monochromatic radiation perturbatively, and it only yields radiation components whose frequencies are integer multiples of $\omega$. 

We will thus show that meson-kink interactions have a very different character in the quantum regime as compared with the classical regime, with the former leading to positive pressure and the second negative pressure.  To some extent this is not surprising, as an initial state consisting of $N$ mesons will yield a number of meson multiplication events proportional to $N$, while the probability of meson fusion will be of order $O(N^2)$.  Thus one expects meson fusion to dominate for sufficiently intense meson sources.

We begin in Sec.~\ref{revsez} with a review of the linearized kink perturbation theory of Refs.~\cite{mekink, me2loop}.  This quantum field theoretic approach is much more economical than the traditional collective coordinate approach of Refs.~\cite{gjscc,gj76}, in particular in the one-kink sector.  Next in Sec.~\ref{moltsez} we calculate the probability of meson multiplication in a general (1+1)d scalar field theory.  In Sec.~\ref{exsez} we apply this formula to two reflectionless kinks: the sine-Gordon soliton and the $\phi^4$ kink.  As a result of integrability, of course, this process does not occur in the sine-Gordon case.  In Sec.~\ref{numsez}, we numerically evaluate various probabilities associated with meson multiplication in the $\phi^4$ model, such as probability densities and recoil probabilities.  \corr{Finally in Sec.~\ref{ifsez} we address quantum corrections to the initial and final states, which are necessary for them to travel without evolving when far from the kink.  We find that these do not contribute to the meson multiplication probability at the order computed.}

\section{Review} \label{revsez}

We will consider a 1+1d quantum field theory of a Schrodinger picture scalar field $\phi(x)$ and its conjugate $\pi(x)$, defined by the Hamiltonian
\begin{equation}
H=\int d x: \mathcal{H}(x):_a, \quad \mathcal{H}(x)=\frac{\pi^2(x)}{2}+\frac{\left(\partial_x \phi(x)\right)^2}{2}+\frac{V(\sqrt{\lambda} \phi(x))}{\lambda}.
\end{equation}
Here $\lambda$ is a coupling constant.  We consider a potential $V$ with degenerate minima, so that the classical equations of motion have a kink solution $\phi(x,t)=f(x)$.  Here $::_a$ is the usual normal ordering at the mass scale $m$, defined by
\beq
m^2=V^{(2)}(\sqrt{\lambda} f(\pm \infty))\hsp
V^{(n)}(\sqrt{\lambda} \phi(x))=\frac{\partial^n V(\sqrt{\lambda} \phi(x))}{(\partial \sqrt{\lambda} \phi(x))^n}.
\eeq
We assume that the two values of the mass, as defined at $x=\infty$ and $x=-\infty$, are equal, as otherwise the vacuum on one side of the kink will be a false vacuum \cite{wstabile}.

As usual, creation operators can be constructed via a plane wave decomposition of the fields.  These create elementary mesons.  Acting them on the vacuum state creates the Fock space of mesons, which we will call the vacuum sector\footnote{\corr{Recall that we are considering theories with at least two vacua, and so at least two vacuum sectors.  When necessary to avoid confusion, we will distinguish between the vacuum sectors corresponding to the vacua at $f(-\infty)$ and $f(\infty)$, which we will call the left and right vacuum sectors.}}.  Similarly, we will construct creation operators which create mesons in the one-kink sector.  Configurations consisting of a single kink plus any number of mesons will be called the one-kink sector.

Consider the unitary displacement operator
\beq
\df={{\rm Exp}}\left[-i\int dx f(x)\pi(x)\right].
\eeq
Acting $\df$ on the vacuum\footnote{\corr{Here we have assumed that the vacuum corresponds to $\phi(x,t)=0$.  More generally, at a vacuum $\phi(x,t)$ is equal to a constant $f$ and one needs to first act with the adjoint of the displacement operator Exp$[-i f \int dx\pi(x) ]$. In general we will leave this correction to the displacement operator implicit, except in Sec.~\ref{ifsez} where we need to distinguish between the two vacua on the two sides of the kink.}}, one arrives at a state in the one-kink sector.  As always, this state can be time-translated using the Hamiltonian $H$.  

Instead of this active transformation point of view, we wish to view $\df$ as a passive transformation of the Hilbert space which preserves the states but transforms the operators.  Let us explain this more precisely.  We refer to the usual representation of the Hilbert space as the {\it{defining frame}}, in which $H$ is the Hamiltonian which generates time translations and whose eigenvalues are energies.  We define the {\it{kink frame}} as follows.  The Dirac ket $|\psi\rangle$ in the kink frame is defined to represent the state $\df|\psi\rangle$ in the defining frame.


Let us try to understand the properties of the kink frame.  First, consider a state represented by the ket $|K\rangle$ in the defining frame.  Then in the kink frame, this state will be represented by the ket $\df^\dag|K\rangle$.  These are two representations of the same state and so clearly they the have the same number of kinks.   Now, if we used the same operator to measure the number of kinks in both frames, then $\df^\dag|K\rangle$ would have one less kink than $|K\rangle$, which is not the case.  Therefore the kink number operator is different in the two frames, in fact the two realizations of the kink number operator are related by conjugation with $\df$, as is the case with all operators.  For example, the Hamiltonian \corr{and momentum operators} in the kink frame \corr{are} the kink Hamiltonian~$H\p$ \corr{and kink momentum $P\p$}
\beq
H\p=\df^\dag H\df\hsp
\corr{P\p=\df^\dag P\df}
. \label{df}
\eeq
To see this, note that if $|K\rangle$ has energy $E_K$, so that
\beq
H|K\rangle=E_K|K\rangle \label{schrodvec}
\eeq
then
\beq
H\p\df^\dag|K\rangle=\df^\dag H|K\rangle=E\df^\dag|K\rangle \label{schrod}
\eeq
and so its eigenvalues yield the correct spectrum.  Similarly, $e^{-iH\p t}$ is the time evolution operator in the kink frame.

The reason that we introduce the kink frame is that, while the defining-frame eigenvalue equation (\ref{schrodvec}) is nonperturbative if $|K\rangle$ is in the one-kink sector, the corresponding kink-frame equation (\ref{schrod}) is perturbative.  Thus, one can solve for kink states $\df^\dag|K\rangle$ using perturbation theory in the kink frame, and then transform the answer back to the defining frame if needed using $\df$.  This has been done to obtain quantum corrections to kink states and masses in Refs.~\cite{mekink,me2loop}.

What is the kink Hamiltonian $H\p$?  Let $Q_n$ be the $n$-loop quantum correction to the kink mass.  Then we may expand $H\p$ into terms $H\p_n$ which have $n$ factors of $\phi(x)$ and $\pi(x)$ when normal-ordered.  One easily finds
\beq
H\p_0=Q_0\hsp H\p_1=0\hsp
H\p_{n>2}=\lambda^{\frac{n}{2}-1}\int dx \frac{V^{(n)}(\sqrt{\lambda} f(x))}{n !}: \phi^n(x):_a.\label{hn}
\eeq

What about $H\p_2$?  This is the most important term, as its eigenstates are the starting points of the perturbative expansion of the entire one-kink sector.  To write it simply, we will need a short digression.

The kink's normal modes $\g(x)$ are the constant frequency solutions of the classical equations of motion corresponding to $H\p_2$
\beq
\V{2}{\g}(x)=\omega^2{\g}(x)+{\g}^{\prime\prime}(x)\hsp \phi(x,t)=e^{-i\omega t}\g(x). \label{sl}
\eeq
There are three kinds of normal mode.  The first is the real zero-mode $\g_B(x)$ which has zero frequency $\omega_B=0$.  Next, there are complex continuum modes $\g_k(x)$ with frequencies $\ok{}=\sqrt{m^2+k^2}$.  Finally, some kinks enjoy discrete, real shape modes $\g_S(x)$ with $0<\omega_S<m$.
We will fix their normalization via the conditions
 $\g^*_k=\g_{-k}$ and 
\beq
\int dx |{\g}_{B}(x)|^2=1,\
\int dx {\g}_{k_1} (x) {\g}^*_{k_2}(x)=2\pi \delta(k_1-k_2),\ 
\int dx {\g}_{S_1}(x){\g}^*_{S_2}(x)=\delta_{S_1S_2}. \label{comp}
\eeq

As $\g(x)$ satisfy a Sturm-Liouville equation (\ref{sl}), they are a complete basis of the space of bounded functions and so can be used to decompose the Schrodinger picture field \cite{cahill76}
\bea
\phi(x) &=&\phi_0 \mathfrak{g}_B(x)+\ppin{k} \left(B_k^{\ddag}+\frac{B_{-k}}{2 \omega_k}\right) \mathfrak{g}_k(x) \label{dec}\\
\pi(x) &=&\pi_0 \mathfrak{g}_B(x)+i \ppin{k}\left(\omega_k B_k^{\ddag}-\frac{B_{-k}}{2}\right) \mathfrak{g}_k(x) \nonumber
\eea
where $B_k^{\ddagger}=B_k^{\dagger} /\left(2 \omega_k\right)$ and $B_{-S}=B_S$.  The symbol $\dint$ is an integral over continuum modes $k$ plus a sum over shape modes $S$.  We have decomposed $\phi(x)$ and $\pi(x)$ into operators $\phi_0,\ \pi_0,\ B$\ and $B^\ddag$ which satisfy the algebra
\beq
\left[\phi_0, \pi_0\right]=i, \quad\left[B_{S_1}, B_{S_2}^{\ddagger}\right]=\delta_{S_1 S_2}, \quad\left[B_{k_1}, B_{k_2}^{\ddagger}\right]=2 \pi \delta\left(k_1-k_2\right).
\eeq

Using this basis, we can write $H\p_2$ as
\begin{equation}
H\p_2=Q_1+H_{\text {free }}, \quad H_{\text {free }}=\frac{\pi_0^2}{2}+ \ppin{k} \omega_k B_k^{\ddag} B_k. \label{h2}
\end{equation}
Now we can interpret the operators.  $\phi_0$ and $\pi_0$ are the position and momentum of a free quantum mechanical particle representing the center of mass of the kink.  The operators $B_S^\ddag$ and $B_k^\ddag$ create bound and continuum normal modes respectively.  The ground state $\vac_0$ of $H\p_2$, which is the kink frame first approximation to the kink ground state $\vac$, is the simultaneous ground state of each of the quantum mechanics terms in Eq.~(\ref{h2}).  Therefore it is the solution of the conditions
\beq
\pi_0\vac_0=B_k\vac_0=B_S\vac_0=0. \label{v0}
\eeq
A general one-meson, one-kink state is, at this leading order, $|k\rangle_0=B^\ddag_k\vac_0$ while acting on this with $B^\ddag_{k\p}$ yields a two-meson, one-kink state 
\beq
|kk\p\rangle_0=B^\ddag_{k}B^\ddag_{k\p}\vac_0. \label{2m}
\eeq

\corr{What has become of translation invariance?  In the defining frame, the time translations are generated by $H$ and spatial translations by $P$.  These commute, and states such as the ground kink and its Fock space excitations are simultaneous eigenvectors of both.  In the kink frame, they are generated by $H\p$ and $P\p$.   A quick calculation, using the sign convention
\beq
\g_B(x)=-\frac{f\p(x)}{\sqrt{Q_0}}
\eeq
yields
\beq
P\p=P+\sqrt{Q_0}\pi_0\hsp 
P=-\int dx \pi(x)\partial_x \phi(x).
\eeq
Intuitively, $P$ is the meson momentum and $\sqrt{Q_0}\pi_0$ is the kink momentum.  These are not separately conserved in the kink frame.  However $P\p$ and $H\p$ commute and all of the simultaneous eigenstates of $P$ and $H$ in the defining frame are also simultaneous eigenstates of $P\p$ and $H\p$ when written in the kink frame, with the same eigenvalues.}

\corr{Linearized kink perturbation theory can be applied to localized wave packets\footnote{\corr{Here we are discussing wave packets in which the center of mass of the kink-meson system is localized, breaking the rigid translation symmetry, which simultaneously displaces the kink and the mesons.  These are distinct from the meson wave packets that we will use below, in which the distance between the meson and kink is localized but the states are invariant with respect to the rigid translation operator $P\p$.}}, where it has been used to compute quantum corrections to form factors in Refs.~\cite{meff,hengyuan}.  In the case of integrable models, these corrections reduce to the known results of Refs.~\cite{weisz77,bab01}. } 

\corr{However in the present paper, we will instead be in interested exclusively in translation-invariant states.  These are states which, in the kink frame, are annihilated by $P\p$.  The fact that $P\p$ yields zero, and not a constant, means that we work in the center of mass frame.  The kink ground state, for example, is translation-invariant.  These translation-invariant states are the quantum field theory analogues of constant wave functions in quantum mechanics\footnote{\corr{Like them, these states are not normalizable.  Below we will see that the same norms appear in the numerator and denominator of various expressions and so will naively cancel them.  In a companion paper \cite{menorm} we introduce an infrared regularization scheme and show that this cancellation does not lead to corrections at the order considered here.}}.  Constant wave function states in quantum mechanics are infinite superpositions of position eigenstates, with a coefficient that is independent of the particle position.  Similarly, here the translation-invariant states are superpositions of kink-meson complexes, with an infinite sum over the position of the center of mass weighted by a coefficient that is independent of this position.  Therefore the kink and the mesons are equally likely to be anywhere.  However $P\p$ shifts the kink-meson system rigidly, and so the distance between the kink and meson can be localized.
}

\corr{In 1+1 dimensions, massless scalar fields are an obstruction to quantization \cite{colemanm}.  Therefore we will consider only models with $m>0$.  As a result, the force exerted by the kink on the mesons is suppressed exponentially in the distance times $m$.  This means that at separations much larger than $1/m$, the meson and kink contributions to $P\p$ are essentially separately conserved.  Furthermore, at such large separations, the contribution of each meson to the momentum is given simply by $k$, up to corrections of order $O(\lambda)$.  This is not to say that the kink does not affect the mesons at very large distances, but rather to say that a distant kink serves only to shift the values of some translation-invariant meson self-couplings, and it does not cause the mesons to accelerate.} 

\section{Meson Multiplication} \label{moltsez}

\subsection{Gaussian Wave Packets}
Our initial condition will be a meson wave packet centered at $x_0$
\begin{equation}
\Phi(x)=\operatorname{Exp}\left[-\frac{\left(x-x_0\right)^2}{4 \sigma^2}+i x k_0\right], \quad x_0 \ll-\frac{1}{ m}, \quad  \frac{1}{k_0},\frac{1}{m}\ll\sigma \ll\left|x_0\right| .
\end{equation}
The bounds on $x_0$ and $|x_0|$ ensure that the initial wave packet, which starts at $x=x_0$, does not overlap with the kink, which is centered at $x=0$.  The lower bounds on $\sigma$ ensure that the meson momentum is sufficiently strongly peaked that all components move towards the kink and also we can approximate, as described below, the wave packet to be monochromatic.

The evolution of the wave packet will be simpler after a kind of Fourier transform 
\begin{equation}
\Phi(x)=\int \frac{d k}{2 \pi} \alpha_k \mathfrak{g}^*_k(x), \quad \alpha_k=\int d x \Phi(x) \mathfrak{g}_k(x).
\end{equation}
This transform is not with respect to the plane waves, which are solutions of the free equations of motion in the vacuum sector, but rather with respect to the normal modes, which are solutions in the one-kink sector.  The shape modes and zero mode need not be included in the transform, as they have support at $|x|$ of order $O(1/m)$, where $\Phi(x)$ is negligibly small.

The initial one-kink, one-meson state $\left|\Phi\right\rangle_0$ can be constructed, in the kink frame, in terms of the free kink ground state $\vac_0$ as
\begin{equation}
\left|\Phi\right\rangle_0=\int d x \Phi(x)\left|x\right\rangle_0=\int \frac{d k}{2 \pi} \alpha_k\left|k\right\rangle_0, \quad\left|k\right\rangle_0=B_k^{\ddagger}|0\rangle_0, \quad|x\rangle_0=\int \frac{d k}{2 \pi} \mathfrak{g}_{k}(x)\left|k\right\rangle_0.\label{init}
\end{equation}

\corr{Eq.~(\ref{init}) is a choice of initial state.  Our strategy, in this section, will be to simply assume this initial condition and evaluate the probability that the final state is in some similarly arbitrarily-defined subspace of the Hilbert space.  This is well-defined.  However, the claim that this choice of initial state and final states is related to meson multiplication is nontrivial.  In particular, quantum corrections to these initial and final states enter at the same order as the amplitude that we will calculate.  We will discuss these initial and final state corrections in Sec.~\ref{ifsez}.}

\subsection{Time Evolution} \label{evsez}
The interactions in the kink frame are summarized by the Hamiltonian terms in Eq.~(\ref{hn}).  These are organized into a power series in $\sqrt{\lambda}$.  \corr{At order $O(\lambda^0)$, only $\hpt$ contributes to the evolution
\beq
|\Phi(t)\rangle_0 |_{O(\lambda^0)}=e^{-i\hpt t}\left|\Phi\right\rangle_0 = \pin{k} \alpha_k e^{-i\ok{} t}\left|k\right\rangle_0=\int dx \pin{k} \alpha_k e^{-i\ok{} t} \g_{-k}(x) |x\rangle_0 .
\eeq
The coefficient
\beq
\Phi(x,t)=\pin{k} \alpha_k e^{-i\ok{} t} \g_{-k}(x)=\int d y \Phi(y) \pin{k}\mathfrak{g}_k(y) e^{-i\ok{} t} \g_{-k}(x)
\eeq
is, to this order, the profile of the meson wave packet.  We may write it in terms of the propagator $G$ as
\beq
\Phi(x,t)=\int dy \Phi(y)G(x,y,t)\hsp G(x,y,t)=\pin{k}\mathfrak{g}_k(y) e^{-i\ok{} t} \g_{-k}(x).
\eeq
For concreteness, consider a reflectionless kink.  Then we will see below that at $x\ll -1/m$
\beq
\alpha_k \g_{-k}(x)=2\sigma\sqrt{\pi}e^{-\sigma^2\left(k-k_0\right)^2}e^{ik_0x+i(k-k_0)(x-x_0)}.
\eeq
Using the linear expansion of $\ok{}$ at $k\sim k_0$, which will be introduced in Eq.~(\ref{om}), one finds
\bea
\Phi(x,t)&=&2\sigma\sqrt{\pi} e^{ik_0 x-i\ok{0}t}\pin{k}  e^{-i(k-k_0)\frac{k_0 t}{\ok{0}}} e^{-\sigma^2\left(k-k_0\right)^2}e^{i(k-k_0)(x-x_0)}\nonumber\\
&=&e^{ik_0 x-i\ok{0}t}{\rm Exp}\left[ 
-\frac{1}{4\sigma^2}\left(x-x_0-\frac{k_0 t}{\ok 0}\right)^2
\right].
\eea
}

\corr{We thus identify $x_0+k_0 t/\ok 0$ as the position of the leading order part of the localized wave packet at time $t$.  In particular, before nearing the kink, the meson wave packet moves at a constant velocity of $k_0/\ok 0$.  It does not accelerate.}

At the next order, $O(\sqrt{\lambda})$, the only term which contributes to meson multiplication is\footnote{Here we have exchanged the order of the $k$ and $x$ integrals with respect to the definition in Eqs.~(\ref{hn}) and (\ref{dec}).  These integrals do not actually commute, and as a result $V_{-k_1k_2k_3}$ appears to be the integral of a nonintegrable function.  It should therefore be remembered that to make sense of this integral, one needs to perform the $k$ integration first.  It turns out that this is equivalent to first performing the $x$ integration using a principal value prescription which will be defined in Eq.~(\ref{iden}).\label{foot}}
\bea
H_I&=&\frac{\sqrt{\lambda}}{4} \int \frac{d k_1}{2 \pi} \frac{d k_2}{2 \pi} \frac{d k_3}{2 \pi} V_{-k_1 k_2 k_3} \frac{1}{\omega_{k_1}} B_{k_2}^{\ddagger} B_{k_3}^{\ddagger} B_{k_1} \\
V_{-k_1 k_2 k_3}&=&\int d x V^{(3)}(\sqrt{\lambda} f(x)) \mathfrak{g}_{-k_1}(x) \mathfrak{g}_{k_2}(x) \mathfrak{g}_{k_3}(x).\nonumber
\eea
$H_I$ converts a one-meson state into a two-meson state
\begin{equation}
H_I |k_1\rangle_0=\frac{\sqrt{\lambda}}{4 \omega_{k_1}} \int \frac{d k_2}{2 \pi} \frac{d k_3}{2 \pi} V_{-k_1 k_2 k_3}\left|k_2 k_3\right\rangle_0.
\end{equation}

At time $t$,  at order $O(\sqrt{\lambda})$, the wave packet evolves to
\begin{equation}
\begin{aligned}
|\Phi(t)\rangle_0&=e^{-i\left(H_{\text {free }}+H_I\right) t}|_{O(\sqrt{\lambda})}\left|\Phi\right\rangle_0 \\
&=\sum_{n=1}^{\infty} \frac{(-i t)^n}{n !}\left(H_{\text {free }}+H_I\right)^n|_{O(\sqrt{\lambda})}\left|\Phi\right\rangle_0 =\sum_{n=1}^{\infty} \frac{(-i t)^n}{n !} \sum_{m=0}^{n-1} H_{\text {free }}^m H_I H_{\text {free }}^{n-m-1}\left|\Phi\right\rangle_0 \\
&=\int \frac{d k_1}{2\pi} \frac{d k_2}{2\pi} \frac{d k_3}{2 \pi} \frac{\sqrt{\lambda}}{4} \alpha_{k_1} V_{-k_1 k_2 k_3} \sum_{n=1}^{\infty} \frac{(-i t)^n}{n !} \sum_{m=0}^{n-1}\left(\omega_{k_2}+\omega_{k_3}\right)^m \omega_{k_1}^{n-m-2}\left|k_2 k_3\right\rangle_0 \\
&=-\frac{i \sqrt{\lambda}}{4} \int \frac{d k_1}{2 \pi} \frac{d k_2}{2 \pi} \frac{d k_3}{2 \pi} \frac{\alpha_{k_1} }{\omega_{k_1}} V_{-k_1 k_2 k_3} {\rm Exp}\left[-i \frac{\omega_{k_1}+\omega_{k_2}+\omega_{k_3}}{2} t\right] \frac{\sin \left(\frac{\omega_{k_2}+\omega_{k_3}-\omega_{k_1}}{2} t \right)}{\left(\omega_{k_2}+\omega_{k_3}-\omega_{k_1}\right)/2}  \left|k_2 k_3\right\rangle_0. \label{fit}
\end{aligned}
\end{equation}
Here we dropped the $O(\lambda^0)$ term which will not contribute to the matrix elements below.  

One may define the Dirac bra corresponding to a one-kink, two-meson state (\ref{2m}) by
\begin{equation}
{}_0\langle k_2 k_3|= \left(B_{k_2}^{\ddagger} B_{k_3}^{\ddagger}|0\rangle_0\right)^\dag={}_0\langle 0|\frac{B_{k_2}}{2\ok{2}}\frac{B_{k_3}}{2\ok{3}}.
\end{equation}
This leads to the normalization\footnote{\corr{The matrix elements ${}_0\langle k_2 k_3|k_2^{\prime} k_3^{\prime}\rangle_0$ and ${_0}{\langle 0}|0\rangle_0$ are both infinite, however only their ratio will appear in the probability of meson multiplication.  In Ref.~\cite{menorm} we show that, to leading order, this ratio agrees with the naive calculation in Eq.~(\ref{inf}).}}
\begin{equation}
{}_0\langle k_2 k_3|k_2^{\prime} k_3^{\prime}\rangle_0=\frac{{_0}{\langle 0}|0\rangle_0}{4 \omega_{k_2} \omega_{k_3}} \left(2 \pi \delta\left(k_2^{\prime}-k_2\right) 2 \pi \delta\left(k_3^{\prime}-k_3\right)+2 \pi \delta\left(k_2^{\prime}-k_3\right) 2 \pi \delta\left(k_3^{\prime}-k_2\right)\right). \label{inf}
\end{equation}
Our master formula for the unnormalized meson multiplication amplitude is then
\begin{equation}
\frac{{}_0\langle k_2 k_3 | \Phi(t)\rangle_0}{{_0}\langle 0| 0\rangle_0}=-\frac{i \sqrt{\lambda}}{8 \omega_{k_2} \omega_{k_3}} \int \frac{d k_1}{2 \pi}\frac{ \alpha_{k_1} }{\omega_{k_1}} V_{-k_1 k_2 k_3} {\rm Exp}\left[-i \frac{\omega_{k_1}+\omega_{k_2}+\omega_{k_3}}{2} t\right] \frac{\sin \left(\frac{\omega_{k_2}+\omega_{k_3}-\omega_{k_1}}{2} t \right)}{\left(\omega_{k_2}+\omega_{k_3}-\omega_{k_1}\right)/2}  . \label{elt}
\end{equation}

\subsection{Amplitude at Finite Times}

Writing the amplitude as
\beq
{}_0\langle k_2 k_3 | \Phi(t)\rangle_0=\frac{ \sqrt{\lambda}}{8 \omega_{k_2} \omega_{k_3}}  \int \frac{d k_1}{2 \pi}\frac{ \alpha_{k_1} }{\omega_{k_1}} V_{-k_1 k_2 k_3} \frac{e^{-i(\ok{2}+\ok{3}) t }-e^{-i\ok{1}t }}{\left(\omega_{k_2}+\omega_{k_3}-\omega_{k_1}\right)}  {_0}\langle 0| 0\rangle_0 \label{amp}
\eeq
we may factor out an overall phase and constant
\beq
A_{k_2k_3}(t)=\frac{e^{i(\ok{2}+\ok{3}) t }}{{_0}\langle 0| 0\rangle_0}{}_0\langle k_2 k_3 | \Phi(t)\rangle_0 = \frac{ \sqrt{\lambda}}{8 \omega_{k_2} \omega_{k_3}}  \int \frac{d k_1}{2 \pi}\frac{ \alpha_{k_1} }{\omega_{k_1}} V_{-k_1 k_2 k_3} \frac{1-e^{i(\ok{2}+\ok{3}-\ok{1}) t }}{\left(\omega_{k_2}+\omega_{k_3}-\omega_{k_1}\right)}.
\eeq
At $t=0$, the matrix element vanishes as the sine in the numerator of Eq.~(\ref{elt}) vanishes.  Taking the time derivative one finds
\bea
\dot{A}_{k_2k_3}(t)&=& -i\frac{ \sqrt{\lambda}}{8 \omega_{k_2} \omega_{k_3}}  \int \frac{d k_1}{2 \pi}\frac{ \alpha_{k_1} }{\omega_{k_1}} V_{-k_1 k_2 k_3} e^{i(\ok{2}+\ok{3}-\ok{1}) t }.\label{aeq}
\eea
This can be simplified with a few good approximations.  

\subsubsection{Reflectionless Kinks}

First of all, $|x_0|\gg\sigma$ and $|x_0|\gg1/m$ and so the Gaussian factor in $\alpha_{k_1}$ has support in the large $|x|$ region, where $\g^*_{k_1}$ is a sum of plane waves.  Let us first consider the case of a reflectionless kink, in which case
\bea
\g_k(x)&=&\left\{\begin{tabular}{lll}
$\mb_ke^{-ikx}$&\rm{if} & $x\ll  -1/m$\\
$\md_ke^{-ikx}$&\rm{if} & $x\gg 1/m$\\
\end{tabular}
\right. \label{gk}\\
|\mb_k|^2&=&|\md_k|^2=1\hsp
\mb^*_k=\mb_{-k}\hsp
\md^*_k=\md_{-k}\nonumber
\eea
where the phases $\mb_k$ and $\md_k$ vary on scales of order $O(m)$ in $k$-space
\beq
\frac{\partial_k\mb_k}{\mb_k}\sim\frac{\partial_k\md_k}{\md_k}\sim O\left(\frac{1}{m}\right).
\eeq
As $x_0\ll -1/m$, this approximation yields
\beq \label{ak1}
\alpha_{k_1}=2\sigma\sqrt{\pi}\mb_{k_1}e^{-\sigma^2\left(k_1-k_0\right)^2}e^{i(k_0-k_1)x_0}.
\eeq

Next, let us consider $t\gg1/m$.  We will not assume that the time is big enough for the meson to arrive at the kink.  So with this approximation, the process will be roughly on-shell, and so $\ok{1}$ can be replaced with $\ok{2}+\ok{3}$.  This needs to be done delicately, as terms of order $\ok{2}+\ok{3}-\ok{1}$ have appeared in various places.  Each expression should be treated as an expansion in powers of $\ok{2}+\ok{3}-\ok{1}$.  However, this replacement can safely by done on the $\ok{1}$ in the denominator of Eq.~(\ref{aeq}), as this term is of zeroth order in $\ok{2}+\ok{3}-\ok{1}$.  

With these two approximations we find
\bea \label{adot}
\dot{A}_{k_2k_3}(t)&=& -i2\sigma\sqrt{\pi}\frac{ \sqrt{\lambda}}{8 \omega_{k_2} \omega_{k_3}(\ok{2}+\ok{3})}  \pin{k_1}\mb_{k_1}
e^{-\sigma^2\left(k_1-k_0\right)^2} e^{i(k_0-k_1)x_0}\nonumber\\
&&\times\left[ \int d y V^{(3)}(\sqrt{\lambda} f(y)) \mathfrak{g}_{-k_1}(y) \mathfrak{g}_{k_2}(y) \mathfrak{g}_{k_3}(y) \right]e^{i(\ok{2}+\ok{3}-\ok{1}) t }.
\eea
$k_1$ is always close to $k_0$, as $\sigma\gg 1/m$, and so we may expand
\begin{equation}\label{om}
\omega_{k_1}=\omega_{k_0}+\left(k_1-k_0\right) \frac{k_0}{\omega_{k_0}}\hsp \mb_{k_1}=\mb_{k_0}\hsp \g_{-k_1}=\g_{-k_0}.
\end{equation}
Inserting Eq.~(\ref{om}) into Eq.~(\ref{adot}),
\bea
\dot{A}_{k_2k_3}(t)&=& -i2\sigma\sqrt{\pi}\mb_{k_0}\frac{ \sqrt{\lambda}e^{i(\ok{2}+\ok{3}-\ok{0}) t }}{8 \omega_{k_2} \omega_{k_3}(\ok{2}+\ok{3})}  \left[ \int d y V^{(3)}(\sqrt{\lambda} f(y)) \mathfrak{g}_{-k_0}(y) \mathfrak{g}_{k_2}(y) \mathfrak{g}_{k_3}(y) \right]\nonumber\\
&&\times\int \frac{d k_1}{2 \pi}
e^{-\sigma^2\left(k_1-k_0\right)^2} e^{i(k_0-k_1)(x_0+\frac{k_0}{\ok{0}}t)}\nonumber\\
&=&-i\mb_{k_0}\frac{ \sqrt{\lambda}e^{i(\ok{2}+\ok{3}-\ok{0}) t }}{8 \omega_{k_2} \omega_{k_3}(\ok{2}+\ok{3})} {\rm Exp}\left[-\frac{(x_0+\frac{k_0}{\ok{0}}t)^2}{4\sigma^2}\right] V_{-k_0 k_2 k_3}.
\eea
\corr{Note that in replacing $V_{-k_1k_2k_3}$ by $V_{-k_0k_2k_3}$ we have assumed that the $k_1$-dependence of $V$ is on scales much broader than $1/\sigma$.  This assumption breaks down near $k_1+k_2+k_3=0$ if $\V3$ does not have compact support, as $V$ may have a $\delta$ function term and also a pole.  These occur far from the mass shell, and so do not reflect any interesting dynamical processes, but rather are an artifact of the fact that our initial condition (\ref{init}) did not include the quantum corrections necessary to propagate rigidly far from the kink.  We will return to this point in Sec.~\ref{ifsez}.}

\subsubsection{Reflective Kinks}

So far we have only considered reflectionless kinks, such as those of the sine-Gordon and $\phi^4$ models.  However, in general kinks are reflective, and so asymptotically the normal modes are of the form
\bea
\g_k(x)&=&\left\{\begin{tabular}{lll}
$\mb_ke^{-ikx}+\mc_ke^{ikx}$&\rm{if} & $x\ll  -1/m$\\
$\md_ke^{-ikx}+\me_k e^{ikx}$&\rm{if} & $x\gg 1/m$\\
\end{tabular}
\right. \label{gk}\\
|\mb_k|^2+|\mc_k|^2&=&|\md_k|^2+|\me_k|^2=1\hsp
\mb^*_k=\mb_{-k}\hsp
\mc^*_k=\mc_{-k}\hsp
\md^*_k=\md_{-k}\hsp
\me^*_k=\me_{-k}.\nonumber
\eea
Again, our initial wave packet is supported near $x_0\ll-1/m$ and so this approximation allows us to simplify the coefficients $\alpha_{k_1}$
\beq \label{ak1}
\alpha_{k_1}=2\sigma\sqrt{\pi}\left[\mb_{k_1}e^{-\sigma^2\left(k_1-k_0\right)^2}e^{i(k_0-k_1)x_0}+\mc_{k_1}e^{-\sigma^2\left(k_1+k_0\right)^2}e^{i(k_0+k_1)x_0}\right].
\eeq

Substituting this into Eq.~(\ref{aeq}) one finds
\bea
\dot{A}_{k_2k_3}(t)&=& -i2\sigma\sqrt{\pi}\frac{ \sqrt{\lambda}}{8 \omega_{k_2} \omega_{k_3}(\ok{2}+\ok{3})} \int \frac{d k_1}{2 \pi}V_{-k_1 k_2 k_3}e^{i(\ok{2}+\ok{3}-\ok{1}) t }\nonumber\\
&&\times \left[
\mb_{k_1} e^{-\sigma^2\left(k_1-k_0\right)^2} e^{i(k_0-k_1)x_0}+\mc_{k_1} e^{-\sigma^2\left(k_1+k_0\right)^2} e^{i(k_0+k_1)x_0}\right].\label{aref}
\eea

Recall that we have fixed $k_0>0$ so that the wave packet moves to the right, towards the kink.  In the reflectionless case this implied that $k_1>0$.  Now we see that there are two Gaussian factors, the first is supported at $k_1\sim k_0$ but the second is instead supported at $k_1\sim -k_0.$  Thus, while the initial motion of the meson is always to the right, in the reflective case this corresponds to two distinct regions in the one-meson Fock space.

As a result, we will need to consider the expansion of $k_1$ about both $k_0$ and also $-k_0$, which leads to the corresponding expansion for the frequencies
\begin{equation}
\omega_{k_1}=\omega_{k_0}+\left(\pm k_1-k_0\right) \frac{k_0}{\omega_{k_0}}. \label{svil}
\end{equation}

Inserting these two expansions into Eq.~(\ref{aref}), we obtain
\bea
\dot{A}_{k_2k_3}(t)&=& -i2\sigma\sqrt{\pi}\frac{ \sqrt{\lambda}e^{i(\ok{2}+\ok{3}-\ok{0}) t }}{8 \omega_{k_2} \omega_{k_3}(\ok{2}+\ok{3})}
 \int \frac{d k_1}{2 \pi}V_{-k_1 k_2 k_3}
\label{adr}\\
&&\times  \left[\mb_{k_1}
e^{-\sigma^2\left(k_1-k_0\right)^2} e^{i(k_0-k_1)(x_0+\frac{k_0}{\ok{0}}t)}+\mc_{k_1}
e^{-\sigma^2\left(k_1+k_0\right)^2} e^{i(k_1+k_0)(x_0+\frac{k_0}{\ok{0}}t)}\right]\nonumber\\
&=&-i\frac{ \sqrt{\lambda}e^{i(\ok{2}+\ok{3}-\ok{0}) t }}{8 \omega_{k_2} \omega_{k_3}(\ok{2}+\ok{3})} {\rm Exp}\left[-\frac{(x_0+\frac{k_0}{\ok{0}}t)^2}{4\sigma^2}\right]\tilde{V}_{-k_0 k_2 k_3}\nonumber
\eea
where we have defined the shorthand
\beq \label{tildv}
\tilde{V}_{-k_0 k_2 k_3}=\mb_{k_0} V_{-k_0 k_2 k_3}+\mc^*_{k_0} V_{k_0 k_2 k_3}.
\eeq

\subsubsection{Remarks}

As a result of the Gaussian factor, this time derivative of the amplitude is only appreciable when the exponent
\beq
x_t=x_0+\frac{k_0}{\ok{0}}t
\eeq
is small, which occurs at time
\beq
t\sim t_1=  -\frac{\ok{0}}{k_0}x_0
\eeq
when the meson strikes the kink.  

In particular, since $t\geq 0$, we see that this requires $k_0$ and $x_0$ to have opposite signs, which of course is necessary for the meson to move towards the kink.  As $A(0)=0$, we learn that the amplitude $A(t)$ vanishes at $t\ll t_1$, before the collision.

\subsection{Amplitude in the Asymptotic Future}

\subsubsection{The Large Time Limit}

We are interested in the large time limit, when the meson has already scattered with the kink.  At large times $t$ we may integrate Eq.~(\ref{adr}) to obtain
\bea
\stackrel{\rm{lim}}{{}_{t\rightarrow\infty}}A_{k_2k_3}(t)&=&
-i\frac{ \sqrt{\lambda} \tilde{V}_{-k_0 k_2 k_3}}{8 \omega_{k_2} \omega_{k_3}(\ok{2}+\ok{3})}\int_{-\infty}^{\infty} dt  {\rm Exp}\left[-\frac{(x_0+\frac{k_0}{\ok{0}}t)^2}{4\sigma^2}\right]e^{i(\ok{2}+\ok{3}-\ok{0}) t }\nonumber\\
&=&-i\frac{ \sqrt{\lambda} \tilde{V}_{-k_0 k_2 k_3}}{4\omega_{k_2} \omega_{k_3}(\ok{2}+\ok{3})}\sigma\sqrt{\pi}\frac{\ok{0}}{k_0}\nonumber\\
&&\times{\rm{Exp}}
\left[-\sigma^2\frac{\ok{0}^2}{k^2_0}\left(\ok{2}+\ok{3}-\ok{0}\right)^2-i\left(\ok{2}+\ok{3}-\ok{0}\right)\frac{\ok{0}}{k_0}x_0
\right].
\eea
Therefore
\beq
\stackrel{\rm{lim}}{{}_{t\rightarrow\infty}}\frac{\left| {}_0\langle k_2 k_3 | \Phi(t)\rangle_0\right|^2}{|{}_0\langle 0\vac_0|^2}=
\frac{ \pi\lambda\sigma^2 \left|\tilde{V}_{-k_0 k_2 k_3}\right|^2}{16\omega^2_{k_2} \omega^2_{k_3}(\ok{2}+\ok{3})^2}\left(\frac{\ok{0}}{k_0}
\right)^2{\rm{Exp}}
\left[-2\sigma^2\frac{\ok{0}^2}{k^2_0}\left(\ok{2}+\ok{3}-\ok{0}\right)^2
\right]. \label{lim}
\eeq

Let us define the on-shell initial momentum $k_I$ by
\beq\label{I23}
 k_I \equiv  \sqrt{\left(\ok{2}+\ok{3}\right)^2-m^2}
\eeq
so that $\ok{I}=\ok{2}+\ok{3}.$  The Gaussian factor in Eq.~(\ref{lim}) has support at $\ok{0}\sim\ok{I}$.  Therefore, as $k_0$ and $k_I$ are both defined to be positive, in the region in $k_2-k_3$-space with the largest contribution to the probability, $k_0\sim k_I$.  We thus expand
\beq
k_0=k_I+(k_0-k_I)
\eeq
and keep only the leading nonvanishing term in each expression.  This yields
\beq
\stackrel{\rm{lim}}{{}_{t\rightarrow\infty}}\frac{\left| {}_0\langle k_2 k_3 | \Phi(t)\rangle_0\right|^2}{|{}_0\langle 0\vac_0|^2}=
\frac{ \pi\lambda\sigma^2 \left|\tilde{V}_{-k_I k_2 k_3}\right|^2}{16\omega^2_{k_2} \omega^2_{k_3}k_I^2}{\rm{Exp}}
\left[-2\sigma^2\frac{\ok{I}^2}{k^2_I}\left(\ok{I}-\ok{0}\right)^2
\right].
\eeq
Using the same expansion as in Eq.~(\ref{svil}) this simplifies further to 
\beq
\stackrel{\rm{lim}}{{}_{t\rightarrow\infty}}\frac{\left| {}_0\langle k_2 k_3 | \Phi(t)\rangle_0\right|^2}{|{}_0\langle 0\vac_0|^2}=
\frac{ \pi\lambda\sigma^2 \left|\tilde{V}_{-k_I k_2 k_3}\right|^2}{16\omega^2_{k_2} \omega^2_{k_3}k_I^2}e^{
-2\sigma^2\left(k_{I}-k_{0}\right)^2
}.
\eeq

\subsubsection{A Faster Derivation}

A faster approach, which however sheds no light on the evolution at intermediate times, is to directly take the $t\rightarrow\infty$ limit of Eq.~(\ref{elt}).  Using the identity
\beq
\stackrel{\rm{lim}}{{}_{t\rightarrow\infty}}
\frac{\sin \left(\frac{\omega_{k_2}+\omega_{k_3}-\omega_{k_1}}{2} t \right)}{\left(\omega_{k_2}+\omega_{k_3}-\omega_{k_1}\right)/2} 
=2 \pi \delta\left(\omega_{k_2}+\omega_{k_3}-\omega_{k_1}\right)=\frac{\omega_{k_I}}{k_I}\left(2 \pi \delta\left(k_1-k_I\right)+2 \pi \delta\left(k_1+k_I\right)\right)
\eeq
the amplitude can be simplified to 
\begin{equation}
\stackrel{\rm{lim}}{{}_{t\rightarrow\infty}}
\frac{{}_0\langle k_2 k_3 | \Phi(t)\rangle_0}{{_0}\langle 0| 0\rangle_0}=-\frac{i \sqrt{\lambda}}{8 \omega_{k_2} \omega_{k_3} k_I}  e^{-i \omega_{k_I} t}\left(\alpha_{k_I} V_{-k_I k_2 k_3}+\alpha_{-k_I} V_{k_I k_2 k_3}\right).
\end{equation}
As $k_I$ and $k_0$ are both large and positive, the Gaussians in Eq.~(\ref{ak1}) with $(k_I+k_0)$ are exponentially suppressed, leaving only the $\mb_{k_I}$ term in $\alpha_{k_I}$ and the $\mc^*_{k_I}$ term in $\alpha_{-k_I}$.  Altogether we find
\beq
\stackrel{\rm{lim}}{{}_{t\rightarrow\infty}}
\frac{{}_0\langle k_2 k_3 | \Phi(t)\rangle_0}{{_0}\langle 0| 0\rangle_0}=-\frac{i\sigma \sqrt{\pi\lambda}}{4 \omega_{k_2} \omega_{k_3} k_I}  e^{-i \omega_{k_I} t}e^{-\sigma^2(k_0-k_I)^2}\tilde{V}_{-k_I k_2 k_3} \label{ampf}
\eeq
in agreement with the longer derivation above.

\subsection{The Probability}

The probability $P$ that $|\Phi(t)\rangle_0$, the state at time $t$, is in a given subspace of the Hilbert space is given by
\begin{equation}
P=\frac{{}_0\langle \Phi(t)|\mathcal{P}|  \Phi(t)\rangle_0}{{}_0\langle \Phi(t) |  \Phi(t)\rangle_0}\label{pdef}
\end{equation}
where $\mathcal{P}$ is a projector onto that subspace.

We are interested in the probability $P_{\rm{tot}}$ that the final state has two mesons, corresponding to the projector 
\begin{equation}
\mathcal{P}_{\rm{tot}}|k_2 k_3\rangle_0=|k_2 k_3\rangle_0\hsp
k_2,\ k_3\in \R.
\end{equation}
We are also interested in the corresponding probability density $P_{\rm{diff}}(k_2,k_3)$ that the final mesons have momenta $k_2$ and $k_3$.  This is related to the total probability by
\beq
P_{\rm{tot}}=\frac{1}{2}\int dk_2 dk_3 P_\text{diff}(k_2,k_3)
\eeq
where the factor of $1/2$ results from the fact that $|k_2k_3\rangle$ and $|k_3k_2\rangle$ represent the same state.  $P_{\rm{diff}}$ is defined by a formula similar to (\ref{pdef}) in which the operator $\mathcal{P}_{\rm{diff}}$ annihilates all states with $k$ not equal to $k_2$ and $k_3$.  It is not a projector, as it has an infinite eigenvalue.  These two equations are easily solved, yielding the operators
\beq
\mathcal{P}_\text{diff}(k_2,k_3)=\frac{\omega_{k_2} \omega_{k_3}}{\pi^2{_0}\langle 0 |0\rangle_0}|k_2 k_3\rangle_0{}_0\langle k_2 k_3|\hsp\mathcal{P}_\text{tot}=\frac{1}{\langle 0 |0\rangle_0}\int d k_2 d k_3\frac{\ok2\ok3}{2\pi^2{_0}} |k_2 k_3\rangle_0{}_0\langle k_2 k_3|.\label{pta}
\eeq

Consider a general reflective kink with $\alpha_{k_1}$ of the form of Eq.~(\ref{ak1})
\begin{equation}
{}_0\langle \Phi(t) |  \Phi(t)\rangle_0={}_0\langle \Phi |  \Phi \rangle_0 =\pin{k_1} \alpha_{k_1} \alpha_{k_1}^{*} \frac{{_0}\langle 0 |0\rangle_0}{2 \omega_{k_1}} =\sqrt{2\pi}\sigma\frac{{_0}\langle 0 |0\rangle_0}{2 \omega_{k_0}}
\end{equation}
where we used $\ok{1}\sim\ok{0}$.

The probability density at a large time $t$ is
\bea
P_{\rm{diff}}(k_2,k_3)&=&\stackrel{\rm{lim}}{{}_{t\rightarrow\infty}}\frac{{}_0\langle \Phi(t)|\mathcal{P}_\text{diff}(k_2,k_3) | \Phi(t)\rangle_0}{{}_0\langle \Phi(t) |  \Phi(t)\rangle_0} =\stackrel{\rm{lim}}{{}_{t\rightarrow\infty}}\frac{\sqrt{2} \ok{0}\ok{2}\ok{3}}{\pi^{5/2}\sigma} \frac{\left| {}_0\langle k_2 k_3 | \Phi(t)\rangle_0\right|^2}{|{}_0\langle 0\vac_0|^2}\nonumber\\
&=&\frac{\lambda\sigma\ok{0} \left|\tilde{V}_{-k_I k_2 k_3}\right|^2}{8\sqrt{2}\pi^{3/2}\omega_{k_2} \omega_{k_3}k_I^2}e^{
-2\sigma^2\left(k_{I}-k_{0}\right)^2
}. \label{pd}
\eea
Note that, by definition, the continuum modes have $k$ real and so this equation only holds when $\ok2,\ok3\geq m$.  Integrating this yields the total probability of meson multiplication at a large time $t$ 
\begin{equation} 
P_{\rm{tot}}=\frac{1}{2}\int dk_2 dk_3 P_{\rm{diff}}(k_2,k_3)=
\frac{ \lambda\sigma\ok{0} }{16\sqrt{2}\pi^{3/2} }
\int  dk_2 dk_3
\frac{  \left|\tilde{V}_{-k_I k_2 k_3}\right|^2}{\omega_{k_2} \omega_{k_3}k_I^2}e^{-2\sigma^2\left(k_{I}-k_{0}\right)^2}.
\end{equation}
As $\sigma\gg 1/m$ we may approximate the Gaussian to be a Dirac delta function, yielding
\bea 
P_{\rm{diff}}(k_2,k_3)&=&\frac{\lambda\ok{I} \left|\tilde{V}_{-k_I k_2 k_3}\right|^2}{16\pi\omega_{k_2} \omega_{k_3}k_I^2}\delta(k_I-k_0)
\label{ptoteq}\\
P_{\rm{tot}}&=&\frac{\lambda\ok{0} }{32\pi k_0^2}
\int dk_2 dk_3
\frac{  \left|\tilde{V}_{-k_I k_2 k_3}\right|^2}{\omega_{k_2} \omega_{k_3}}\delta(k_I-k_0)\nonumber\\
&&\hspace{-2cm}=\frac{ \lambda }{32\pi k_0}
\int_{-\sqrt{(\ok{0}-m)^2-m^2}}^{\sqrt{(\ok{0}-m)^2-m^2}} dk_2
\frac{  \left|\tilde{V}_{-k_0, k_2, \sqrt{(\ok{0}-\ok{2})^2-m^2}}\right|^2+\left|\tilde{V}_{-k_0, k_2, -\sqrt{(\ok{0}-\ok{2})^2-m^2}}\right|^2}{\omega_{k_2} \sqrt{(\ok{0}-\ok{2})^2-m^2}}\nonumber
\eea
where we used
\beq
\frac{\partial k_I}{\partial k_3}=\frac{\ok{0}k_3}{k_0\ok{3}}=\frac{\ok{0}\sqrt{(\ok{0}-\ok{2})^2-m^2}}{k_0(\ok{0}-\ok{2})}.
\eeq


\section{Examples: The Sine-Gordon Soliton and $\phi^4$ Kink} \label{exsez}

\subsection{The Sine-Gordon Soliton}
In the sine-Gordon theory, defined by
\beq
V(\sqrt{\lambda}\phi(x))=m^2\left(1-{\rm{cos}}(\sqrt{\lambda}\phi(x)\right)
\eeq
the symbol $V_{k_1k_2k_3}$ is given in Ref.~\cite{me2loop}
\bea
V_{k_1k_2k_3}&=&\frac{\pi i\sqrt{\lambda}}{4}{\rm{sign}}(k_1k_2k_3){\rm{sech}}\left(\frac{\pi(k_1+k_2+k_3)}{2m}\right)\\
&&\times\frac{(\ok{1}+\ok{2}+\ok{3})(\ok{1}+\ok{2}-\ok{3})(\ok{1}+\ok{3}-\ok{2})(\ok{2}+\ok{3}-\ok{1})}{\ok{1}\ok{2}\ok{3}}.\nonumber
\eea
As a result
\beq
V_{\pm k_Ik_2k_3}=0
\eeq
because it is proportional to $\ok{2}+\ok{3}-\ok{I}=0$.  This in turn implies that
\beq
\tilde{V}_{- k_Ik_2k_3}=0
\eeq
as it is a linear combination (\ref{tildv}) of $V_{\pm k_Ik_2k_3}$.  Eq.~(\ref{pd}) then implies that the differential probability vanishes for all $k_2$ and $k_3$.

This is to be expected, the integrability of the sine-Gordon model implies that the number of mesons is conserved and so meson multiplication does not appear in the $S$-matrix.


\subsection{The $\phi^4$ Kink}

\subsubsection{Review}

We will need an expression for $\tilde{V}_{-k_1k_2k_3}$ in the case of the $\phi^4$ double-well model, with potential
\beq
V(\sqrt{\lambda}\phi(x))=\frac{\lambda\phi^2(x)}{4}\left(\sqrt{\lambda}\phi(x)-\sqrt{2}m\right)^2
.
\eeq
This requires a knowledge of $\mb_k,\ \mc_k$\ and $V_{k_1k_2k_3}$.  The first two are easily read off of the normal modes
\beq
\g_k(x)=\frac{e^{-ikx}}{\ok{} \sqrt{k^2+\b^2}}\left[k^2-2\b^2+3\b^2\sech^2(\b x)-3i\b k\tanh(\b x)\right]\hsp\b=\frac{m}{2}. \label{norm}
\eeq

At $x\ll-1/\beta$ this becomes a plane wave with phase
\beq \label{coeffbc}
\mb_k=\frac{k^2-2\beta^2+3i\beta k}{\ok{}\sqrt{k^2+\beta^2}}\hsp \mc_k=0.
\eeq
As the $\phi^4$ kink is reflectionless, the product $\mb_k\mc_k$ vanishes \cite{merif}.  

Using Eq.~(\ref{tildv}) and $|\mb_k|=1$, the reflectionless condition thus leads to the simplification
\beq 
\left|\tilde{V}_{-k_0 k_2 k_3}\right|=\left|V_{-k_0 k_2 k_3}\right|.
\eeq
We then need only calculate $V_{k_1k_2k_3}$.  In Ref.~\cite{phi42loop} this is calculated in terms of a sum of integrals over $x$, however those integrals are not evaluated because that paper was concerned with infrared divergences which required a delicate treatment of the integrand.  We will see a similar infrared divergence here, arising from the fact that the 3-point interaction responsible for meson multiplication has a nonzero constant norm even far from the kink.  Meson multiplication far from the kink is suppressed only because the corresponding matrix element oscillates quickly, leading to destructive interference when the initial momentum is integrated over even a very small interval.

Let us begin by reviewing the expression for $V_{k_1k_2k_3}$ in Ref.~\cite{phi42loop}.  First, the third derivative of the potential is 
\beq
V^{(3)}(\sqrt{\lambda}f(x))=6\sqrt{2}\b \tanh(\b x).
\eeq
Note that it is of order $O(\sqrt{\lambda})$, and so that will be the order of our amplitude.  Also notice that it tends to a nonzero constant at large $x$ and $-x$.

We will perform the $x$-integrals using the identities
\bea
\int dx e^{-ikx}\sech^{2n}(\b x)&=&\left\{
\begin{array}{cl}
2\pi\delta(k) &  {\rm{\ \ \ if}}\  n=0 \\ \frac{\pi}{(2n-1)!k}\left[\prod_{j=0}^{n-1}\left(\frac{k^2}{\b^2}+(2j)^2\right)\right]\ck   & {\rm{\ \ \ if}}\ n>0
\end{array}
\right.\nonumber\\
\int dx e^{-ikx}\sech^{2n}(\b x)\tanh(\b x)&=&-i\frac{\pi}{(2n)!\b}\left[\prod_{j=0}^{n-1}\left(\frac{k^2}{\b^2}+(2j)^2\right)\right]\ck \label{iden}.
\eea
Note that in the $n=0$ cases of the two integrals, the integrand does not become small at large $|x|$.  These formulas correspond to a kind of principal value prescription for evaluating the integrals.  We have checked that this principal value prescription is indeed the right one, as it yields the same answer as would be achieved by integrating over a small region in $k_1$ with a smooth weight function.  Such a coherent integral was indeed present in our master formula (\ref{elt}) for the amplitude, it is the integral over the momentum in the initial wave packet.  The fact that the $k$ integral should be performed before the $x$ integral was explained in Footnote~\ref{foot}.

$V_{k_1k_2k_3}$ consists of a sum of terms which are each integrals over $x$ of $\sech^{2I}(\beta x)\tanh^J(\beta x)$ where $I\in\{0,1,2,3\}$ and $J\in\{0,1\}$.  The case $I=J=0$ yields a $\delta(k)$, where we have defined
\beq
k=k_1+k_2+k_3.
\eeq
As $\ok{I}=\ok{2}+\ok{3}$, $k$ is not zero and so this term vanishes.  We will keep it, as our expression for $V_{k_1k_2k_3}$ may be useful for future problems, however we will separate it as it will not contribute to meson multiplication at tree level.  Thus we decompose
\beq
V_{k_1k_2k_3}=V^{00}_{k_1k_2k_3}+\hat{V}_{k_1k_2k_3}\hsp
V^{00}_{k_1k_2k_3}=-\frac{9\sqrt{2}i\beta^2 k_1k_2k_3\left(6\b^2+k_{1}^2+k_2^2+k_{3}^2\right)2\pi\delta(k)}{\ok1\ok2\ok3\sqrt{\b^2+k_1^2}\sqrt{\b^2+k_2^2}\sqrt{\b^2+k_3^2}}
\eeq
where $V^{00}$ contains all of the $\delta(k)$ terms and only $\hat{V}$ will be relevant below.

Let us define the symbols $u$ by
\beq
\hat{V}_{k_1k_2k_3}=\frac{6\sqrt{2}\pi\b\ck}{\ok1\ok2\ok3\sqrt{\b^2+k_1^2}\sqrt{\b^2+k_2^2}\sqrt{\b^2+k_3^2}}\sum_{J=0}^1\sum_{I=1-J}^3 u_{k_1k_2k_3}^{IJ}
\eeq
where the sum does not include $I=J=0$, as that term is in $V^{00}$.  

Each $u^{IJ}$ is defined to be the term in $V_{k_1k_2k_3}$ with an $x$ integral of $e^{ixk}\sech^{2I}(\b x)\tanh^J(\b x)$.  Let us define the symbol $\Phi$ to summarize the coefficients
\beq
u_{k_1k_2k_3}^{IJ}=\frac{\sinh\left(\frac{\pi k}{2\beta}\right)}{\pi}\Phi_{k_1k_2k_3}^{IJ}\int dxe^{-ixk}\sech^{2I}(\b x)\tanh^J(\b x).
\eeq
Ref.~\cite{phi42loop} provided the components of $\Phi$  
\bea
\Phi_{k_1k_2k_3}^{10}&=&3i\b\left[16\b^4S_1^1+\b^2\left(-5S_2^{21}-18S_3^1\right)+S_3^1S_2^1\right]\\
\Phi_{k_1k_2k_3}^{20}&=&9i\b^3\left[-7\b^2S^1_1+S_2^{21}+3S_3^1\right]\hsp \Phi_{k_1k_2k_3}^{30}=27i\b^5S_1^1\nonumber\\
\Phi_{k_1k_2k_3}^{01}&=&-8\b^6+\b^4(18S_2^1+4S_1^2)+\b^2(-2S_2^2-9S_3^1S_1^1)+S_3^2
\nonumber\\
\Phi_{k_1k_2k_3}^{11}&=&3\b^2\left[12\b^4+\b^2(-15S_2^1-4S_1^2)+(S_2^2+3S_3^1S_1^1)\right]
\nonumber\\
\Phi_{k_1k_2k_3}^{21}&=&9\b^4\left[-6\b^2+(3S_2^1+S_1^2)\right]
\hsp
\Phi_{k_1k_2k_3}^{31}=27\b^6
\nonumber
\eea
in terms of symmetric combinations of the $k$'s
\bea
S_1^n&=&k_1^n+k_2^n+k_3^n\hsp 
S_2^n=(k_1k_2)^n+(k_1k_3)^n+(k_2k_3)^n\hsp
S_3^n=(k_1k_2k_3)^n\nonumber\\
S_2^{mn}&=&k_1^mk_2^n+k_1^mk_3^n+k_2^mk_3^n+k_1^nk_2^m+k_1^nk_3^m+k_2^nk_3^m.
\eea

\subsubsection{The Calculation}

We may now perform the $x$ integrals using Eq.~(\ref{iden}) 
\bea
u_{k_1k_2k_3}^{I0}&=&\Phi_{k_1k_2k_3}^{I0}\frac{1}{(2I-1)!k}\left[\prod_{j=0}^{I-1}\left(\frac{k^2}{\b^2}+(2j)^2\right)\right]\\
u_{k_1k_2k_3}^{I1}&=&\Phi_{k_1k_2k_3}^{I1}\frac{-i}{(2I)!\b}\left[\prod_{j=0}^{I-1}\left(\frac{k^2}{\b^2}+(2j)^2\right)\right].\nonumber
\eea
In particular, we find
\bea
u_{k_1k_2k_3}^{10}&=&3ik\left[16\b^3S_1^1+\b\left(-5S_2^{21}-18S_3^1\right)+\frac{1}{\beta}S_3^1S_2^1\right]\\
u_{k_1k_2k_3}^{20}&=&\frac{3ik}{2}\left(\frac{k^2}{\beta^2}+4\right)\left[-7\b^3 S^1_1+\b S_2^{21}+3\b S_3^1\right]\nonumber\\
u_{k_1k_2k_3}^{30}&=&\frac{9i k}{40}\left(\frac{k^4}{\beta^4}+20\frac{k^2}{\beta^2}+64\right)\left[\beta^3S_1^1\right]\nonumber\\
u_{k_1k_2k_3}^{01}&=&i\left[8\b^5+\b^3(-18S_2^1-4S_1^2)+\b^1(2S_2^2+9S_3^1S_1^1)-\frac{S_3^2}{\b}\right]
\nonumber\\
u_{k_1k_2k_3}^{11}&=&\frac{3ik^2}{2}\left[-12\b^3+\b(15S_2^1+4S_1^2)+\frac{1}{\b}(-S_2^2-3S_3^1S_1^1)\right]\nonumber\\
u_{k_1k_2k_3}^{21}&=&\frac{3ik^2}{8}\left(\frac{k^2}{\beta^2}+4\right)\left[6\b^3+\b(-3S_2^1-S_1^2)\right]\nonumber\\
u_{k_1k_2k_3}^{31}&=&-\frac{3ik^2}{80}\left(\frac{k^4}{\beta^4}+20\frac{k^2}{\beta^2}+64\right)\b^3.\nonumber
\eea

Reassembling these components, we finally arrive at
\bea \label{vphi4}
\hat{V}_{k_1k_2k_3}
&=&\frac{6\sqrt{2} \pi \csch\left(\frac{\pi (k_1+k_2+k_3)}{2 \b}\right)}{\ok1\ok2\ok3\sqrt{\b^2+k_1^2}\sqrt{\b^2+k_2^2}\sqrt{\b^2+k_3^2}}\nonumber\\
&&\times \Bigg\{8i\b^6  + 5i \b^4 (k_1^2+k_2^2+k_3^2)+2i \b^2  (k_1^2 k_2^2+k_1^2 k_3^2+k_2^2 k_3^2)\nonumber\\
&&\quad +i\left[\frac{3}{16}(-k_1^6-k_2^6-k_3^6+k_1^4 k_2^2+k_1^4 k_3^2+k_2^4 k_3^2\right.\nonumber\\
&&\left.\quad\qquad+k_2^4 k_1^2+k_3^4 k_1^2+k_3^4 k_2^2)+\frac{1}{8}k_1^2k_2^2k_3^2\right]\Bigg\}.
\eea
Recall that the meson multiplication probability density (\ref{pd}) and total probability (\ref{ptoteq}) only require the special case $k_1=-k_I$.  In this case the coefficients simplify to
\bea \label{vphi4I23}
V_{-k_I k_2 k_3}&=&-\frac{48\sqrt{2}\pi i \ok2\ok3\ok{I}\csch\left(\frac{\pi \left(k_2+k_3-k_I\right)}{m}\right)}{\sqrt{4k_2^2+m^2}\sqrt{4k_3^2+m^2}\sqrt{4k_I^2+m^2 }}.
\eea



For completeness we provide $\tilde{V}$
\bea \label{vtildephi4}
\tilde{V}_{-k_I k_2 k_3}&=&\mb_{k_I} V_{-k_I k_2 k_3}+\mc_{-k_I} V_{k_I k_2 k_3}
=\frac{k_I^2-2\beta^2+3i\beta k_I}{\ok{I}\sqrt{k_I^2+\beta^2}}V_{-k_I k_2 k_3}\nonumber\\
&=&\frac{48\sqrt{2}\pi\ok2 \ok3 \left(i \left(3m^2-2\ok{I}^2\right)+3m k_I\right)\csch\left(\frac{\pi \left(k_2+k_3-k_I\right)}{m}\right)}{\sqrt{4k_2^2+m^2}\sqrt{4k_3^2+m^2}\left(4k_I^2+m^2\right)}
\eea
where we used Eq.~(\ref{coeffbc}) and Eq.~(\ref{I23}).  However, as a result of $(\ref{tildv})$, at tree level we only need the absolute value $|\tilde{V}|$ which is equal to $|\hat{V}|$ for a reflectionless kink and to $|V|$ at $k_1\sim - k_I$.

Substituting Eq.~(\ref{vtildephi4}) into  Eq.~(\ref{pd}), we find the probability density and total probability for meson multiplication. Our main result is the following analytic expression for the probability density
\bea
P_{\rm{diff}}(k_2,k_3)&=&\frac{288\sqrt{2\pi} \lambda\sigma \ok0\ok2\ok3\ok{I}^2\csch^2\left(\frac{\pi \left(k_2+k_3-k_I\right)}{m}\right)}{k_I^2(4k_2^2+m^2)(4k_3^2+m^2)(4k_I^2+m^2 )}e^{
-2\sigma^2\left(k_{I}-k_{0}\right)^2}. \label{dpn}
\eea
In the limit $\sigma\rightarrow\infty$ of a monochromatic initial meson this yields
\bea 
P_{\rm{diff}}(k_2,k_3)&=&\frac{\lambda\ok{I} \left|\tilde{V}_{-k_I k_2 k_3}\right|^2}{16\pi\omega_{k_2} \omega_{k_3}k_I^2}\delta(k_I-k_0)\label{princ}\\
&=&\frac{288\pi \lambda \ok2\ok3\ok{I}^3\csch^2\left(\frac{\pi \left(k_2+k_3-k_I\right)}{m}\right)}{k_I^2(4k_2^2+m^2)(4k_3^2+m^2)(4k_I^2+m^2 )}\delta(k_I-k_0). \nonumber
\eea
As expected, it is order $O(\lambda)$.  The Dirac $\delta$ function imposes exact energy conservation.  On the other hand, momentum conservation among mesons is imposed by the csch.  This is not a $\delta$ function, and so the momentum can be transferred between the mesons and the kink.  Note that the condition that $k_2$ and $k_3$ be real implies that this equation is only valid when
\beq
m\leq \ok 2,\ \ok 3\leq \ok 0-m.
\eeq

Integrating over $k_3$, one arrives at the probability density
\bea
P_{\rm{diff}}(k_2)&=&\int dk_3 P_{\rm{diff}}(k_2,k_3)\\
&=&  \frac{ 288\pi\lambda\ok2\ok{0}^2(\ok0-\ok 2)^2}{k_0(4k_2^2+m^2)(4(\ok{0}-\ok{2})^2-3m^2)(4k_0^2+m^2 )\sqrt{(\ok{0}-\ok{2})^2-m^2}}\nonumber\\
&&\times\left[\csch^2\left(\frac{\pi \left(k_2+\sqrt{(\ok{0}-\ok{2})^2-m^2}-k_0\right)}{m}\right)\right.\nonumber\\
&&\left.+\csch^2\left(\frac{\pi \left(k_2-\sqrt{(\ok{0}-\ok{2})^2-m^2}-k_0\right)}{m}\right)\right].
\nonumber
\eea
The last term in the denominator leads to a pole at the threshold $k_3=0$, corresponding to the fact that the Jacobian factor $dk_3/dk_2$ diverges.   At finite $\sigma$ this pole is smeared out. The two csch terms correspond to the $k_3$ travelling in the direction of the original meson or bouncing back, and their arguments are the momentum transfer between the mesons and the kink.  

In the ultrarelativistic limit $k_0\gg m$, Eq.~(\ref{princ}) becomes
\bea
P_{\rm{diff}}(k_2,k_3)
&=&\frac{9\pi \lambda  \csch^2\left(\frac{\pi m}{2k_2k_3k_I}\left(k_I^2-k_2k_3 \right)\right)}{2  k_2 k_3 k_I}\delta(k_I-k_0)\\
&=&\frac{18 \lambda k_2k_3k_0}{\pi m^2\left(k_0^2-k_2k_3 \right)^2}\delta(k_2+k_3-k_0).\nonumber
\eea
This is supported when $k_2,\ k_3$\ and $k_I$ are all of order $k_0$, and so it is proportional to $1/k_0$.  To obtain the total probability, one integrates over the $k_2-k_3$ plane, or more precisely the line $k_2+k_3=k_0$ with $k_2,\ k_3>0$.  The length of this line is of order $O(k_0)$, and so the total probability asymptotes to a constant at large $k_0$.   Letting $k_2=k_0 x$ we find that in the ultrarelativistic limit
\beq
P_{\rm{tot}}
=\frac{9\lambda}{\pi m^2} \int_0^{1} dx \frac{  x (1-x)}{\left(1-x+x^2 \right)^2}
=\frac{\lambda}{m^2} \left(\frac{6}{\pi}-\frac{2}{\sqrt{3}}  \right)\sim 0.755 \frac{\lambda}{m^2}. \label{asy}
\eeq

\section{Numerical Results for the $\phi^4$ Kink} \label{numsez}
In this section we will numerically evaluate some of the probabilities just calculated for the $\phi^4$ double-well model.

At order $O(\lambda)$ the probability density $P_{\rm{diff}}$ and the total probability $P_{\rm{tot}}$ are proportional to $\lambda$, so in the plots we will divide them by $\lambda$. We use the parameters $m=1$, $\sigma=20$. We have numerically checked that as long as the value of $\sigma$ satisfies $1/m\ll\sigma$
, the value of $\sigma$ will not affect the numerical results.

\begin{figure}[htbp]
\centering
\includegraphics[width = 0.6\textwidth]{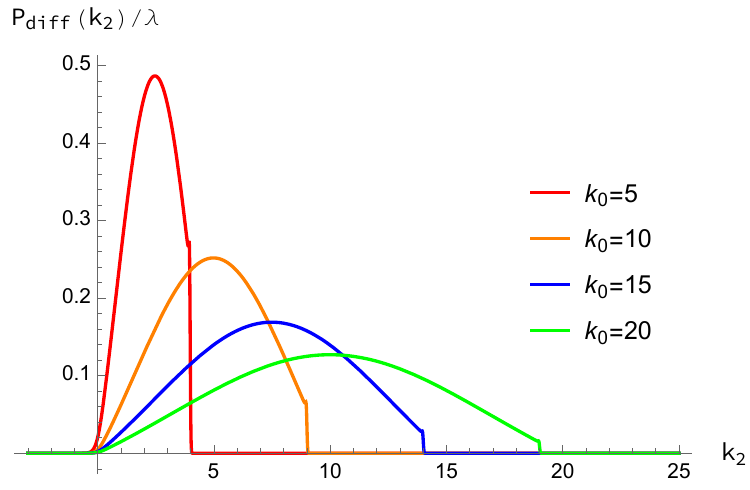}
\caption{The probability density, $P_{\rm{diff}}(k_2)$, that one of the final mesons has momentum $k_2$, plotted for various values of $k_0$.  The factor of $\lambda$ has been divided out.}\label{pdiff}
\end{figure}

We begin in Fig.~\ref{pdiff} by plotting the probability density $P_{\rm{diff}}(k_2)=\int dk_3 P_{\rm{diff}}(k_2,k_3)$, where $P_{\rm{diff}}(k_2,k_3)$ is taken from Eq.~(\ref{dpn}), that one of the two final mesons will have momentum $k_2$.  The shoulder on the right of each curve is not a numerical artifact.  It results from the fact that, with fixed $k_0$, the Jacobian factor in the $k_3$ integral diverges at threshold for the production of the corresponding meson.  This would lead to a pole in the limit $\sigma\rightarrow\infty$, but here this pole is smeared by the momentum width of the initial wave packet.

\begin{figure}[htbp]
\centering
\includegraphics[width = 0.6\textwidth]{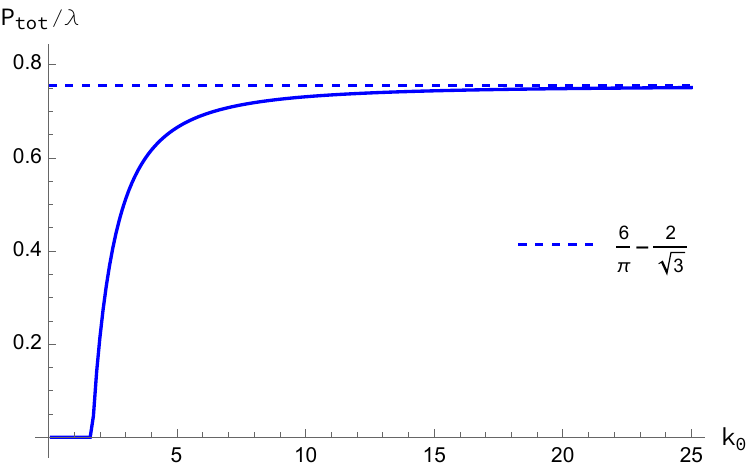}
\caption{The total meson multiplication probability $P_{\rm{tot}}$ as a function of $k_0$, rescaled by $1/\lambda$.  The dashed line is the asymptotic value derived in Eq.~(\ref{asy}).}\label{ptot}
\end{figure}

Next, in Fig.~\ref{ptot}, we plot the total probability for meson multiplication, as a function of the initial meson momentum $k_0$.  Note that, at high $k_0$, the probability asymptotes to the value found in Eq.~(\ref{asy}).

\begin{figure}[htbp]
\centering
\includegraphics[width = 0.6\textwidth]{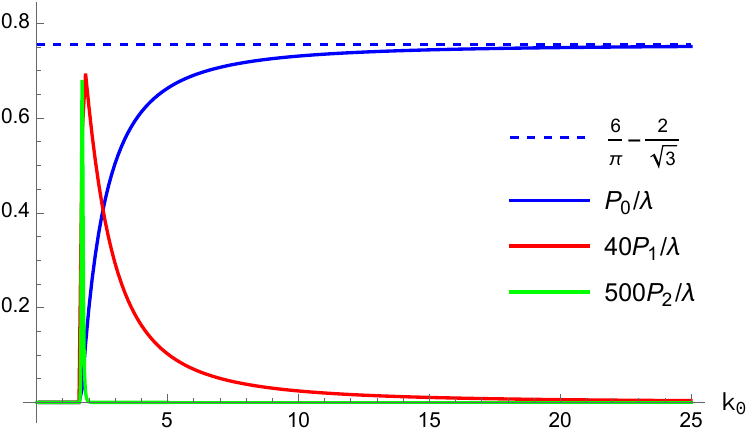}
\caption{The probability $P_n$ that $n$ of the momenta of the outgoing mesons are negative. These are all rescaled by $1/\lambda$ and also by other factors, given in the legend, to make them visible in the plot.  The dashed line is again the asymptotic value in Eq.~(\ref{asy}).}\label{p0p1p2}
\end{figure}

Finally in Fig.~\ref{p0p1p2} we plot the probability, $P_n$, that precisely $n$ of the final mesons have $k<0$, so that they travel backwards from the kink.  This plot shows that, at order $O(\lambda)$, even reflectionless kinks lead to some reflection.  However, as might be expected, this is very rare when the momentum $k_0$ of the initial meson is much greater than the meson mass $m$.

\section{\corr{Initial and Final States}} \label{ifsez}

\corr{In this section we will try to understand the choice of initial and final states.}

\corr{
\subsection{Corrections to the Amplitude}
The amplitude that we calculated (\ref{ampf}) is of order $O(\sl)$.  It results from the product of an initial wave function, a term in $H\p_3$ and a final wave function which are, respectively, $O(\lambda^0)$,\ $O(\sl)$ and $O(\lambda^0)$.  However contributions at the same order arrive from corrections to the initial or final state of order $O(\sl)$, so long as one uses the free Hamiltonian $H\p_2$, which is order $O(\lambda^0)$.}

\corr{The free Hamiltonian conserves the meson quantum numbers, and so these contributions arise from quantum corrections to the initial state containing 2-mesons and corrections to the final state containing 1-meson.  The  calculations in the previous sections correspond to simply setting such corrections to zero.  While such an initial condition is allowed, and such a definition of a two-meson state can be forced, both are unnatural as they are not eigenstates of the interacting Hamiltonians relevant to any regime in this problem.  In fact, if $O(\sl)$ corrections are not included in the initial state, then they will be dynamically generated, oscillating as the wave packet propagates.  However, this oscillation at subleading order does not affect the conclusion that the wave packet does not accelerate far from the kink, nor does it affect the leading order probability of meson multiplication.}

\corr{In this section we will make another, better motivated, choice of quantum corrections to the initial and final states, and show that it does not affect our amplitude at the order calculated.}

\corr{
\subsection{Constructing the Initial State}
The initial state is a one-meson wave packet which approaches the kink from the left.  To the left of the kink, the classical kink solution $f(x)$ approaches a minimum $f_L=f(-\infty)$ of the potential $V(\sl f(x))$.}  

\corr{\subsubsection{Outline of the Construction}
}

\corr{At times $t<0$, the meson has always been in the left vacuum, and has never been close to the kink.  Therefore, we want to construct an initial state, at time $t=0$, such that the meson wave packet is a nearly monochromatic superposition of eigenstates not of the full kink Hamiltonian $H\p$, but rather of the left vacuum Hamiltonian $H_L$
\beq
H_L=\mathcal{D}_L^\dag H\mathcal{D}_L\hsp
\mathcal{D}_L={\rm{Exp}}\left[-i\int dx f_L\pi(x) 
\right]. \label{hl}
\eeq}

\corr{The left vacuum evolution operator $e^{-iH_Lt}$ acts on our meson wave packet constructed from $H_L$ eigenstates by rigidly translating it, with no acceleration or deformation apart from the usual smearing.  As the two Hamiltonians $H_L$  and $H\p$ act identically on mesons far to the left of the kink, the true evolution operator $e^{-iH\p t}$ also acts on the meson wave packet by rigid translation, without acceleration or deformation, before it approaches the kink.   Thus this construction will define a suitable one-meson, one-kink asymptotic state to set up our scattering problem.  As this wave packet is not an eigenstate of the kink Hamiltonian, evolving it forward in time, it will evolve nontrivially once the meson wave packet reaches the kink.}

\corr{To construct this state is easy.  In Ref.~\cite{menormal} one-meson Hamiltonian eigenstates were constructed for a sector described by an arbitrary classical solution $f(x)$.  While in most applications, $f(x)$ is taken to be in the kink sector, the derivation in fact works for any static, classical solution $f(x)$ in any sector.  In particular it applies equally well to the left vacuum solution $f(x)=f_L$ or to the right vacuum solution $f(x)=f_R$.  One may follow all of the arguments of Ref.~\cite{menormal} simply replacing $f(x)$ by $f_L$ to obtain one-meson states in the left vacuum, in the left vacuum frame.  The active transformation $\df\mathcal{D}_L^\dag$ adds a kink at the origin, leading to a one-meson one-kink state, while staying in the left vacuum frame.   One then performs the passive transformation $\mathcal{D}_L\df^\dag$ which leaves the state unchanged but changes the frame of the Hilbert space from the left vacuum frame to the kink frame.  Putting these two transformations together we find that, with no transformation at all, one may directly interpret the so-constructed one-meson states in the left vacuum frame as one-meson, one-kink states in the kink frame.  Of course the former were eigenstates of the left vacuum frame evolution operator $H_L$ while the latter are not eigenstates of the kink frame evolution operator $H\p$, which is the reason that we get any dynamics at all.
}

\corr{The $H_L$ eigenstates that we have constructed are quite different from $H\p$ eigenstates.  However, they can be assembled into localized wave packets at $x\ll 0$ and, so long as they remain localized at $x\ll 0$, they will behave as free particles because the left vacuum and kink Hamiltonians will act on them identically.  Indeed, at $x\ll 0$, the difference between the kink Hamiltonian $H\p$ and the left vacuum Hamiltonian is exponentially suppressed in $m|x|$.}

\corr{Such wave packets have three properties which make them suitable as initial conditions.  First, they have been defined using the Hamiltonian $H_L$ with no kink, as expected for a meson wave packet that has not yet interacted with the kink.  Second, as we will show below, under evolution using the kink Hamiltonian $H\p$ they propagate via rigid translations, with constant velocity and no deformation, before they arrive at the kink.  Finally, at leading order they are our old wave packets (\ref{init}) from Sec.~\ref{moltsez}. Recall that the old wave packet (\ref{init}) only evolved under $e^{-iH\p t}$ via rigid translations at leading order, whereas at order $O(\sl)$ it was deformed as it evolved.}

\corr{\subsubsection{Explicit Construction}
}

\corr{Let us look at the leading order corrections explicitly.  Any state $|\psi\rangle$ may be expanded as
\beq
|\psi\rangle=\sum_{mn}\phi_0^m \ppink{n}\gamma^{mn}_\psi (k_1\cdots k_n)|k_1\cdots k_n\rangle_0
\eeq
for some coefficient functions $\gamma_\psi$.  Then the leading order term in the one-kink, one-meson state $|\kt\rangle$, used throughout this paper, was
\beq
\gamma_{\kt}^{01}(k_1)=2\pi\delta(k_1-\kt).
\eeq
In a general sector, there are a number of next order corrections.  However, in a vacuum sector, the normal modes reduce to plane waves.  For simplicity, let us consider a reflectionless kink so that these plane waves can be identified with the continuum normal modes on the far left $\g_k(x)=\mb_k e^{-ikx}$, where $\mb_k$ is a phase \cite{merif}.  Then all of the corrections in Ref.~\cite{menormal} vanish except for two
\bea
\gamma_{\kt}^{02}(k_1,k_2)&=& \frac{\sqrt{\lambda}V^{(3)}(\sl f_L) \mb_{k_1}\mb_{k_2}\mb_{-k_1-k_2} 2\pi \delta({k_1+k_2-\kt})}{4\omega_\kt\left(\omega_\kt-\ok1-\ok2\right)}\\
\gamma_{\kt}^{04}(k_1\cdots k_4)&=&-\frac{\sqrt{\lambda}V^{(3)}(\sl f_L)  \mb_{k_1}\mb_{k_2}\mb_{k_3}2\pi \delta({k_1+k_2+k_3})}{6\sum_{j=1}^3 \ok{j}}2\pi\delta(k_4-\kt).\nonumber
\eea
These contributions to the one-meson state both arise from the three-meson vertex.  The first arises when the vertex converts one meson into two, the second when it creates three mesons while leaving the already existing meson alone.
}

\corr{
In summary, we propose that the bare $|k_1\rangle_0$ be replaced by
\bea
|k_1\rangle_L&=&|k_1\rangle_0+\frac{\sqrt{\lambda}V^{(3)}(\sl f_L) }{4\ok1}\pin{k_2}\frac{\mb_{k_1-k_2}\mb_{k_2}\mb_{-k_1}|k_2,k_1-k_2\rangle_0 }{\ok1-\ok2-\omega_{k_1-k_2}}\nonumber\\
&&-\frac{\sqrt{\lambda}V^{(3)}(\sl f_L) }{6}\pin{k_2}\pin{k_3}\frac{\mb_{k_2}\mb_{k_3}\mb_{-k_2-k_3}|k_1,k_2,k_3,-k_2-k_3\rangle_0}{\ok2+\ok3+\omega_{k_2+k_3}} \label{k1c}
\eea
in the construction of the initial state (\ref{init}).  These are the order $O(\sl)$ corrections, which are the only ones relevant to the $O(\sl)$ amplitude treated in this note.  At higher orders, the corrections are again derived as in Ref.~\cite{menormal}, with $f(x)$ replaced by $f_L$.
}

\corr{
\subsection{Early Time Evolution of the Initial State}
We have proposed the initial state
\beq 
|\Phi\rangle_L=\pin{k_1}\alpha_{k_1} |k_1\rangle_L. \label{init2}
\eeq
Our claim is that at times well before the collision,  this new initial state, unlike (\ref{init}), evolves under the full kink Hamiltonian by a simple displacement at a constant velocity, and so it is suitable for a conventional scattering interpretation of our process.  Let us now show that this is the case.
}

\corr{
For brevity, we will ignore the four-meson part of the state, since it can be treated identically to the two-meson part by including, in $H_I$, the term in $H_3$ with three $B^\ddag$ operators.  Now we want to find the order $O(\sl)$ contributions to $e^{-iH\p t}|\Phi\rangle_L$.  Most of these were already found and reported in Eq.~(\ref{fit}).  The only new terms arise from the free evolution of the two-meson correction to the state
\bea
e^{-i\hpt t}\left(|\Phi\rangle_L-|\Phi\rangle_0\right)&=&e^{-i\hpt t}\pink{2}\alpha_{k_1}\frac{\sqrt{\lambda}V^{(3)}(\sl f_L) }{4\ok1}\frac{\mb_{k_1-k_2}\mb_{k_2}\mb_{-k_1}|k_2,k_1-k_2\rangle_0 }{\ok1-\ok2-\omega_{k_1-k_2}}\nonumber\\
&&\hspace{-3cm}=\frac{\sqrt{\lambda}V^{(3)}(\sl f_L)}{4} \pink{2}\frac{\alpha_{k_1}}{\ok1}\frac{e^{-it(\ok2+\omega_{k_1-k_2})}\mb_{k_1-k_2}\mb_{k_2}\mb_{-k_1}|k_2,k_1-k_2\rangle_0 }{\ok1-\ok2-\omega_{k_1-k_2}}.\label{hec}
\eea
}

\corr{This term may be rewritten as
\bea
|\Phi(t)\rangle_L&=&e^{-iH\p t}|\Phi\rangle_L\supset e^{-i\hpt t}\left(|\Phi\rangle_L-|\Phi\rangle_0\right)=\pin{k_1}e^{-i\ok 1 t}\alpha_{k_1} \left(|k_1\rangle_L-|k_1\rangle_0\right) \label{sup}\\
&&\hspace{-2cm}+\frac{\sqrt{\lambda}V^{(3)}(\sl f_L)}{4} \pink{2}\frac{\alpha_{k_1}}{\ok1}\frac{\left(e^{-it(\ok2+\omega_{k_1-k_2})}-e^{-i\ok1 t}\right)\mb_{k_1-k_2}\mb_{k_2}\mb_{-k_1}|k_2,k_1-k_2\rangle_0 }{\ok1-\ok2-\omega_{k_1-k_2}}. \nonumber
\eea
The first term on the right hand side corresponds to rigid motion without deformation, what about the second?  To obtain the total evolution $|\Phi(t)\rangle_L$, one needs to add the contributions in Eq.~(\ref{fit}).  Adding the correction on the second line of (\ref{sup}) to (\ref{fit}), one finds that (\ref{fit}) is modified via the replacement
\beq
V_{k_1k_2k_3}\rightarrow V_{k_1k_2k_3}-V^{(3)}(\sl f_L)\mb_{k_1}\mb_{k_2}\mb_{k_3}2\pi\delta(k_1+k_2+k_3).
\eeq
This replacement in the 3-meson interaction $V_{k_1k_2k_3}$ exactly removes the contribution to the evolution from the only interaction in the left vacuum that is present at this order: the momentum-conserving three-meson vertex.}

\corr{Now recall that we have argued in Subsec.~\ref{evsez} that, except for the $\delta(-k_1+k_2+k_3)$ term in $V_{-k_1k_2k_3}$, the amplitude does not evolve before the meson reaches the kink.  Now, as promised, we have tied up this loose end: the apparent evolution in (\ref{fit}) at $k_1=k_2+k_3$ is canceled by the evolution (\ref{hec}) of the higher order correction (\ref{k1c}) to the initial condition (\ref{init2}).   Only the first line in (\ref{sup}) is not canceled by (\ref{fit}).  As a result, when folded into a wave function that has support at $x\ll 0$, the corrected state $|k_1\rangle$ evolves as $e^{-i\ok{1} t}|k_1\rangle$ under the full kink Hamiltonian evolution operator $e^{-iH\p t}$, as claimed.}  

\corr{This is in accord with the physical picture proposed above.  Indeed, at $x_0\ll 0$, the one-meson to two-meson process can only occur at $k_1+k_2+k_3=0$ because the kink is too far to exchange momentum with the mesons.  Thus the meson system itself conserves momentum at these early times. 
}

\corr{What have we gained?  We see that our initial wave packet $|\Phi\rangle_L$ has a well-defined and constant momentum in the asymptotic past, and its profile including its leading quantum correction remains unchanged before the meson wave packet arrives at the kink.  The simple phase rotation (\ref{sup}) at each $k_1$ corresponds, via the same standard arguments used in Subsec.~\ref{evsez}, into the rigid motion of a wave packet with momentum centered at $k_0$, up to the usual spreading effects.  In particular, although the presence of the kink affects the meson self-interactions even at an infinite distance, these interactions are translation-invariant.  Indeed, they are those of the vacuum sectors.  Therefore there is no long distance acceleration, which would have implied that the usual scattering matrix is ill-defined~\cite{dollard,morchio}.  In such a case, the kink would have been able to affect the meson at a distance, leading to a memory effect \cite{stromem} and in particular long-distance information in the states \cite{faddress,other}.}

\corr{\subsection{Final State Corrections}
We have argued that the one-meson states $|k_1\rangle_0$ that we used to construct our initial wave packet in Eq.~(\ref{init}) are not ideal choices, because a quantum correction of order $O(\sl)$ will be generated well before reaching the kink.  We found a prescription for a quantum correction to the initial state which makes it travel unperturbed until it reaches the kink.  We called the quantum corrected initial state $|k_1\rangle_L$.}

\corr{The probability is determined by the initial conditions, the Hamiltonian and the projector onto the final states that would trigger the detector.  We have considered quantum corrections to the first two.  In this subsection we will consider quantum corrections to the projector.  In Sec.~\ref{moltsez} we considered the uncorrected projector (\ref{pta}).  More generally, if $|\alpha\rangle$ is an orthonormal basis of a subspace of the Hilbert space, then the projector
\beq
\mathcal{P}=\int d\alpha |\alpha\rangle\langle \alpha|
\eeq
yields, when sandwiched between the a state and itself, the probability that the state is in the subspace spanned by the states $|\alpha\rangle$.}


\corr{
Here $\alpha$ is an abstract index on the basis $|\alpha\rangle$ of final states that trigger the detector.  What properties need these states satisfy?   In principle, any choice corresponds to some detector and so leads to a well-defined probability.  However, we will define meson multiplication by imposing three conditions on these final states $|\alpha\rangle$.  First, at leading order they should consist of two mesons $|k_2k_3\rangle_0$.  Second, in the far past and future, the action of the projector should be independent of time. In the far past and future, the state is described by a wave packet that is localized far to the left or the right of the kink.  Therefore the projector should be constructed from states which are time-independent on the two sides of the kink.  In other words, these states should be 2-meson states of the Hamiltonian for the vacuum sector on each respective side of the kink.  These Hamiltonians are $H_L$, defined in Eq.~(\ref{hl}), and $H_R$, defined identically but with $f_L$ replaced by $f_R$.}

\corr{But how can a state $|\alpha\rangle$ be constructed of eigenstates for two distinct, non-commuting, Hamiltonians $H_L$ and $H_R$?  
One can construct the projector from a basis of localized wavepacket states which, on the left and right of the kink, are superpositions of eigenstates $|k_1k_2\rangle_L$ and $|k_1k_2\rangle_R$ of the left and right vacuum Hamiltonians respectively.  
}

\corr{
Finally, we demand that the probability of observing a two-meson final state at the beginning of the experiment must be equal to 0.  Thus, we need to choose quantum corrections so that the projector annihilates our initial state.}

\corr{This does not entirely fix the projector, nor the states $|\alpha\rangle$.  However, since we are only searching for the $O(\sl)$ piece of $|\alpha\rangle$, and we are interested in the $O(\sl)$ piece of the matrix element $\langle \alpha|e^{-iH\p t}|k_1\rangle_L$, we need only consider the inner product with the $O(\lambda^0)$ part of $|k_1\rangle_L$, which is $|k_1\rangle_0$.  In general one needs to be careful about contributions from zero modes in such arguments, but in a companion paper \cite{menorm} we find an exact formula for such inner products and 
show that corrections to such naive calculations are nonzero but are suppressed by a power of $O(\sl)$, although they mix sectors whose meson number differs by one.  Thus these corrections do not affect the amplitude at $O(\sl)$.}

\corr{For wave packets localized at $x\ll 0$, in Eq.~(\ref{k1c}) we have required that the leading corrections to $|k_1\rangle_L$ have a certain form.  Let us define another set of states, $|k_1\rangle_R$, which have similar corrections but this time corresponding to the vacuum on the right, where $f_R=f(\infty)$
\bea
|k_1\rangle_R&=&|k_1\rangle_0+\frac{\sqrt{\lambda}V^{(3)}(\sl f_R) }{4\ok1}\pin{k_2}\frac{\md_{k_1-k_2}\md_{k_2}\md_{-k_1}|k_2,k_1-k_2\rangle_0 }{\ok1-\ok2-\omega_{k_1-k_2}}\nonumber\\
&&-\frac{\sqrt{\lambda}V^{(3)}(\sl f_L) }{6}\pin{k_2}\pin{k_3}\frac{\md_{k_2}\md_{k_3}\md_{-k_2-k_3}|k_1,k_2,k_3,-k_2-k_3\rangle_0}{\ok2+\ok3+\omega_{k_2+k_3}}. \label{k1c2}
\eea
Here $\md_k$ are phases such that, at $x\gg 0$, $\g_k(x)=\md_k e^{-ikx}$.}

\corr{The inner product of the corrections $|k_1\rangle_L-|k_1\rangle_0$ and $|k_1\rangle_R-|k_1\rangle_0$, relevant far to the left and right of the kink, with respect to $|k_2k_3\rangle_0$ are
\bea
\frac{{}_0\langle k_2k_3|\left(|k_1\rangle_L-|k_1\rangle_0\right)}{{}_0\langle 0|0\rangle_0}&=&
\frac{\sqrt{\lambda}V^{(3)}(\sl f_L) \mb_{k_2}\mb_{k_3}\mb_{-k_2-k_3}2\pi\delta({k_2+k_3-k_1})}{8\ok{2}\ok{3}\omega_{k_1}\left(\omega_{k_1}-\ok2-\ok3\right)}\\
\frac{{}_0\langle k_2k_3|\left(|k_1\rangle_R-|k_1\rangle_0\right)}{{}_0\langle 0|0\rangle_0}&=&
\frac{\sqrt{\lambda}V^{(3)}(\sl f_R) \md_{k_2}\md_{k_3}\md_{-k_2-k_3}2\pi\delta({k_2+k_3-k_1})}{8\ok{2}\ok{3}\omega_{k_1}\left(\omega_{k_1}-\ok2-\ok3\right)}.\nonumber
\eea
To cancel them, one requires that the corrections to $|k_2k_3\rangle_0$ include
\bea
|k_2k_3\rangle_L&=&|k_2k_3\rangle_0+\frac{\sl V^{(3)}(\sl f_L)\mb_{-k_2}\mb_{-k_3}\mb_{k_2+k_3}}{4\ok2\ok3(\ok2+\ok3-\omega_{k_2+k_3})}|k_2+k_3\rangle_0\label{lr}\\
|k_2k_3\rangle_R&=&|k_2k_3\rangle_0+\frac{\sl V^{(3)}(\sl f_R)\md_{-k_2}\md_{-k_3}\md_{k_2+k_3}}{4\ok2\ok3(\ok2+\ok3-\omega_{k_2+k_3})}|k_2+k_3\rangle_0\nonumber
\eea
where we have used the properties $\mb^*_k=\mb_{-k}$ and $\md^*_k=\md_{-k}$. Corrections to other terms in the $n$-meson Fock space are allowed, but this is the only correction that has nonvanishing inner product with $|k_1\rangle_0$ at this order, and so the only term which can contribute to the final state correction.
}

\corr{Note that the projector $\mathcal{P}$ is not constructed by summing over all $|k_2k_3\rangle_L {}_L\langle k_2k_3|$ and $|k_2k_3\rangle_R {}_R\langle k_2k_3|$.  Rather, it is constructed from a basis of localized wave packets which, when localized at $x\ll 0$, are constructed from $|k_2k_3\rangle_L$ and when localized at $x\gg 0$, are constructed from $|k_2k_3\rangle_R$.   There is no need to include states with meson wave packets localized near the kink, as these will never appear in the asymptotic past or future.  In practice, inner products of these 2-meson states with localized states may, with exponentially-suppressed imprecision, be obtained by simply inserting the formula (\ref{lr}) for $|k_2k_3\rangle_L$ or $|k_2k_3\rangle_R$ depending on where the states are localized.}

\corr{
\subsection{Correction to the Amplitude}
In Sec.~\ref{moltsez} we computed the amplitude
\beq
{}_0\langle k_2k_3|e^{-iH\p t}|k_1\rangle_0. \label{prima}
\eeq
We are now interested in the corrections appearing in
\beq
\frac{{}_L\langle k_2k_3|e^{-i\hpt t}|k_1\rangle_L}{{}_0\langle 0|0\rangle_0}\ \ \rm{and}\ \ \frac{{}_R\langle k_2k_3|e^{-i\hpt t}|k_1\rangle_R}{{}_0\langle 0|0\rangle_0}. \label{esp}
\eeq
The initial and final state corrections to the probability at time $t$ are calculated from matrix elements of wave packets localized near the position $x_0+k_0 t/\ok 0$.  As a result, only the first term in (\ref{esp}) is relevant at early times $t\ll-x_0 \ok 0/k_0$, and only the second at late times $t\gg-x_0 \ok 0/k_0$.}
  
\corr{Assembling the results above, the corresponding initial and final state corrections to the first expression in Eq.~(\ref{esp}) are respectively
\bea
&&\frac{\sl V^{(3)}(\sl f_L)\mb_{k_2}\mb_{k_3}\mb_{-k_2-k_3}}{4\ok2\ok3(\ok2+\ok3-\omega_{k_2+k_3})}{}_0\langle k_2+k_3|e^{-i\hpt t}|k_1\rangle_0\\
&&\hspace{4cm}=
\frac{\sl V^{(3)}(\sl f_L)e^{(-i\omega_{k_2+k_3}t)}\mb_{k_2}\mb_{k_3}\mb_{-k_1}2\pi\delta(k_2+k_3-k_1)}{8\ok2\ok3\omega_{k_2+k_3}(\ok2+\ok3-\omega_{k_2+k_3})}\nonumber\\
&&{}_0\langle k_2k_3|e^{(-i\hpt t)}\frac{\sqrt{\lambda}V^{(3)}(\sl f_L)\mb_{k_2}\mb_{k_3}\mb_{-k_2-k_3} }{4\ok1}\pin{k\p}\frac{|k\p,k_1-k\p\rangle_0 }{\ok1-\okp{}-\omega_{k_1-k\p}}\nonumber\\
&&\hspace{4cm}=
-\frac{\sl V^{(3)}(\sl f_L)e^{(-i\omega_{k_2+k_3}t)}\mb_{k_2}\mb_{k_3}\mb_{-k_1}2\pi\delta(k_2+k_3-k_1)}{8\ok2\ok3\omega_{k_2+k_3}(\ok2+\ok3-\omega_{k_2+k_3})}.\nonumber
\eea
One may observe that these two corrections cancel precisely, and so the meson multiplication probability before the collision is unaffected by initial and final state corrections.  In other words, the probability is still zero.  The calculation proceeds similarly for the second term in (\ref{esp}), using the matrix elements valid on the right side of the kink, and so the meson multiplication probability after the collision is also unaffected by initial and final state corrections.  Roughly speaking, we have shown that (\ref{esp}) and (\ref{prima}) are equal, at order $O(\sl)$.  We conclude that the adiabatic approximation (\ref{init}) yields the correct meson multiplication amplitude at leading order.}  


\corr{This result was obvious from the beginning.  Far from the kink, the mesons conserve momentum and energy among themselves and so meson splitting is kinematically forbidden.  Initial and final state corrections, on the other hand, result from meson splitting and fusion respectively far before or after interacting with the kink.}

\section{Remarks}
Expanding the potential of the $\phi^4$ double-well model about one of its minima, one finds a cubic interaction.  This interaction, in principle, allows a meson to split into two mesons.  However, this process is forbidden in the vacuum because it is not possible to simultaneously conserve energy and momentum.

On the other hand, in the presence of a kink the situation changes.  At leading order in perturbation theory, the mesons still cannot transfer energy to the kink.  However the momentum can be transferred if the meson splits sufficiently close to a kink.  This transfer appears in the probability density (\ref{princ}) as a csch${}^2$ term which enforces approximate momentum conservation among the mesons.

The momentum transfer at a distance nonetheless complicates our calculations, as the meson splitting can occur at any position and all of these positions need to be integrated over, naively leading to these divergences.  We have found three ways of treating these divergences.  First, the coherent integral over the momentum of the initial meson wave packet causes the rapidly oscillating amplitude at large $|x|$ to be suppressed.  Next, adding an exponential damping term to the amplitude and then taking the limit as the damping vanishes also removes the divergence.  Finally, the principal value prescription for the $x$ integral of tanh, used above, renders it finite.  We have checked that all three methods of removing the divergence yield the same results.  Only the first is justified, as it results from the intrinsic spread of the wave packet and not an {\it{ad hoc}} modification.  However the later two methods are much more easily implemented in our calculations.

There are only two inelastic processes that may occur in the scattering of a kink with a single meson at order $O(\lambda)$.  One is meson splitting, treated here.  The second is the (de)excitation of a shape mode while the meson is transmitted or reflected.  We intend to turn to this process in the near future.

\section* {Acknowledgement}

\noindent
JE is supported by NSFC MianShang grants 11875296 and 11675223. HL acknowledges the support from CAS-DAAD Joint Fellowship Programme for Doctoral students of UCAS.

\end{document}